\documentclass[aps,noeprint,onecolumn,nofootinbib,notitlepage,superscriptaddress]{revtex4-2}
\usepackage{graphicx}
\usepackage{dcolumn}
\usepackage{bm}
\usepackage{amssymb}
\usepackage{float}
\usepackage{multirow}
\usepackage{threeparttable}
\usepackage{slashed}
\usepackage{xcolor}
\usepackage{soul}
\usepackage{braket}
\usepackage{physics}
\usepackage{multirow}
\usepackage{gensymb}
\usepackage[colorlinks=true, pdfstartview=FitV, bookmarks=true, bookmarksnumbered=true, breaklinks]{hyperref}
\usepackage{mathtools,braket}
\usepackage[normalem]{ulem}
\usepackage{textgreek}
\usepackage{mathrsfs}
\usepackage{booktabs}
\usepackage{etoolbox}
\usepackage{amsmath}
\numberwithin{equation}{section}
\renewcommand{\theequation}{\arabic{section}.\arabic{equation}}

\usepackage{lipsum}
\usepackage{color}
\definecolor{blue}{rgb}{0.0, 0.0, 1.0}
\definecolor{red}{rgb}{1.0, 0.0, 0.0}
\definecolor{royalblue}{rgb}{0.0, 0.14, 0.4}
\hypersetup{colorlinks=true, linkcolor=black, citecolor=blue, urlcolor=blue}

\newcommand{\Nraw}{N_0}
\newcommand{\Npv}{\mathcal N_{\mathrm{PV}}}

\newcommand{\be}{\begin{eqnarray}}
\newcommand{\ee}{\end{eqnarray}}
\newcommand{\ba}{\begin{array}}
\newcommand{\ea}{\end{array}}
\newcommand{\bea}{\begin{eqnarray}}
\newcommand{\eea
}{\end{eqnarray}}
\newcommand{\bi}{\begin{itemize}}
\newcommand{\ei}{\end{itemize}}

\newcommand{\nc}{\providecommand}
\nc{\non}{\nonumber}
\nc{\hc}{\hbox {H.c.}}
\nc{\noi}{\noindent}
\nc{\barx}{\bar{x}}
\nc{\pbarn}{\;\hbox {pb}}
\nc{\fbarn}{\;\hbox {fb}}
\nc{\cone}{{\scriptsize \textcolor{orange}{\circled{1}}}}
\nc{\ctwo}{{\scriptsize \textcolor{orange}{\circled{2}}}}
\nc{\cthr}{{\scriptsize \textcolor{orange}{\circled{3}}}}

\nc{\hsp}{\hspace{0.5cm}}
\nc{\lsp}{\hspace{1cm}}
\nc{\Lsp}{\hspace{2cm}}
\nc{\LLsp}{\lsp\lsp}
\nc{\lra}{\longrightarrow}
\nc{\p}{\prime}
\nc{\sgn}{\text{sgn}}
\nc{\ph}{\varphi}
\nc{\eff}{\mathrm{eff}}
\nc{\sqM}{\sqrt{M}}
\nc{\NL}{\mathrm{NL}}
\nc{\moc}{\mathcal{M}} 
\nc{\rd}[1]{\textcolor{red}{#1}}
\nc{\bl}[1]{\textcolor{blue}{#1}}

\nc{\pbar}{\bar{\psi}}
\nc{\beq}{\begin{equation}}  \nc{\eeq}{\end{equation}}
\nc{\bea}{\begin{eqnarray}}  \nc{\eea}{\end{eqnarray}}
\nc{\baa}{\begin{array}}     \nc{\eaa}{\end{array}}
\nc{\bit}{\begin{itemize}}   \nc{\eit}{\end{itemize}}
\nc{\ben}{\begin{enumerate}} \nc{\een}{\end{enumerate}}
\nc{\bce}{\begin{center}}    \nc{\ece}{\end{center}}
\nc{\bpm}{\begin{pmatrix}}   \nc{\epm}{\end{pmatrix}}
\nc{\bvt}{\begin{verbatim}}  \nc{\evt}{\end{verbatim}}
\nc{\vp}[1]{\mathbf{p}_{#1}}
\nc{\vk}{{\bm k}_\perp}
\nc{\vkp}{{\bm k}'_\perp}
\nc{\vq}{{\bm q}_\perp}
\nc{\uu}{{\uparrow\uparrow}}
\nc{\ud}{{\uparrow\downarrow}}
\nc{\du}{{\downarrow\uparrow}}

\def\lsim{\mathrel{\raise.3ex\hbox{$<$\kern-.75em\lower1ex\hbox{$\sim$}}}}
\def\gsim{\mathrel{\raise.3ex\hbox{$>$\kern-.75em\lower1ex\hbox{$\sim$}}}}

\def\udots{\mathinner{\mkern1mu\raise1pt\vbox{\kern7pt\hbox{.}}\mkern2mu\raise4pt\hbox{.}\mkern2mu\raise7pt\hbox{.}\mkern1mu}}

\def\vp{{\bf p}}

\def\be{\begin{equation}}
\def\ee{\end{equation}}
\def\bea{\begin{eqnarray}}
\def\eea{\end{eqnarray}}

\def\vp{{\bf p}}

\def\be{\begin{equation}}
\def\ee{\end{equation}}
\def\bea{\begin{eqnarray}}
\def\eea{\end{eqnarray}}

\def\orcid#1{\kern .08em\href{https://orcid.org/#1}{\includegraphics[keepaspectratio,width=0.7em]{ORCID_iD.png}}}

\newcommand{\safeincludegraphics}[2][]{%
  \IfFileExists{#2}{\includegraphics[#1]{#2}}{%
    \fbox{\parbox[c][2.3in][c]{5.65in}{\centering Missing figure file:\\[0.4em]
      \texttt{\detokenize{#2}}}}%
  }%
}

\begin{document}

\title{Bakamjian--Thomas construction and light-front zero modes in pseudoscalar transitions: A comparison with the covariant Bethe--Salpeter model}

\author{Ho-Meoyng Choi}
\email{homyoung@knu.ac.kr}
\affiliation{Department of Physics Education, Teachers College,
Kyungpook National University, Daegu 41566, Republic of Korea}

\begin{abstract}
We examine the role of light-front zero modes in the extraction
of the semileptonic form factor $f_-(q^2)$ between pseudoscalar
mesons. In an exactly solvable covariant Bethe--Salpeter (BS)
model, we calculate the transverse-current zero-mode contribution
directly from the shrinking nonvalence region and reproduce it
with an effective operator in the valence convolution. In the
valence-only light-front quark model (LFQM) based on the
Bakamjian--Thomas (BT) construction, the one-body current is
built from on-shell constituent spinors, and the interaction
enters through the mass operator and wave functions. The
$(+,\perp)$ extraction of $f_-$ is zero-mode free and is
saturated by the on-shell overlap. Using the free constituent
invariant masses in the internal minus-current projection gives
the same $f_-$ from $(+,-)$ after integration. For spacelike momentum transfer, we prove this
equality for rotationally invariant radial wave functions with
the light-front Jacobian and the adopted pseudoscalar spin
coupling, provided the reduced overlaps converge. Subtracting
the minus-current result obtained with physical meson masses
from the common transverse result gives the BT completion.
Current-component independence indirectly identifies this completion
as the effective zero-mode contribution missing from the physical-mass
minus-current extraction.
When the BS valence current and the effective zero-mode operator
for the nonvalence endpoint contribution are transferred unchanged
to the Gaussian BT-based LFQM, the resulting expression fails to
reproduce the component-independent BT result for $f_-$.
We construct a mass-weighted complex transverse contour on which
the Gaussian wave functions remain analytic for
$0<q^2<(m_1+m_2)^2$, where $m_1$ and $m_2$ are the initial and
final active-quark masses. We illustrate the method by numerically evaluating the
$D\to K$ form factors.
\end{abstract}

\maketitle

\section{Introduction}
\label{sec:introduction}
The extraction of a hadronic form factor tests not only the
bound-state wave function but also its consistent use with the
current operator. In light-front (LF) calculations, this issue is
particularly transparent because different current components can
probe different contributions to the same amplitude.
The central question addressed here is how zero-mode
contributions can be identified and quantified in a covariant
model with an explicit nonvalence sector and in a valence-only
light-front quark model (LFQM). We examine how the
Bakamjian--Thomas (BT) construction~\cite{BT53,Poly10,CCKP} supplies an internal
current-component consistency criterion in the latter.

Pseudoscalar-to-pseudoscalar semileptonic transitions provide a simple
setting for this question, since two form factors suffice and
independent current-component extractions can be compared.
Measured decay distributions probe these form
factors~\cite{BaBar07DK,Belle02BD}, and the form-factor normalization
enters the determination of Cabibbo--Kobayashi--Maskawa (CKM)
matrix elements~\cite{Cabibbo63,KM73,PDG26}. Reliable extractions
and CKM unitarity tests consequently require control of the
hadronic matrix elements~\cite{PDG26,CKMfitter05,UTfit05,FLAG24}.
Here we focus on the internal consistency of the form-factor
calculation rather than on a new phenomenological fit.

The matrix elements are conventionally described by $f_+(q^2)$ and
$f_-(q^2)$, or equivalently by $f_+(q^2)$ and the scalar form factor
$f_0(q^2)$. These form factors have been studied using lattice
QCD~\cite{HPQCD21,FermilabMILC23}, light-cone sum
rules~\cite{Khodjamirian09,BallZwicky05}, Dyson--Schwinger and
Bethe--Salpeter (BS) methods~\cite{Yao20,JiMaris01,Yao22},
relativistic constituent-quark
models~\cite{Faustov20,MelikhovStech00,ScoraIsgur95,Heger21},
and LFQMs~\cite{Jaus90,Jaus91,CCH97,CCH04,CJ99b,CJBc,QC18}.
In the LF approach, boost-invariant wave functions provide a
convenient connection between internal constituent motion and
current matrix elements expressed as
overlaps~\cite{BrodskyPauliPinsky98}.

The simplicity of an overlap representation does not by itself
eliminate instantaneous, nonvalence, or zero-mode
contributions~\cite{BCJ01,BCJ03,MF,deMelo02,CJ98,ChoiJi05ZM}.
When the $q^+=0$ Drell--Yan--West (DYW) frame is approached from
$q^+>0$, a finite contribution can survive as the nonvalence
interval collapses to a longitudinal endpoint. This surviving
contribution is an LF zero mode. The plus current may be zero-mode
free even when the transverse or minus component is not.
An exactly solvable covariant BS model therefore provides a useful
benchmark because its manifestly covariant amplitude, LF valence
contribution, and nonvalence endpoint contribution can be calculated
independently~\cite{BCJ01,BCJ03,MF,Jaus99}. In the unequal-mass BS
model considered here, $f_+$ is saturated by the plus-current
valence contribution, whereas the $(+,\perp)$ current-component
extraction of $f_-$ requires a finite zero mode. The direct
nonvalence calculation and the effective valence-space operator
representation reproduce the same zero-mode contribution within
this model.

The BT construction~\cite{BT53,Poly10,CCKP} provides a different
realization of the interacting meson. The interaction is incorporated
in the invariant mass operator, $\hat M=\hat M_0+\hat V_{q\bar q}$,
while the kinematic LF generators retain their free-particle form.
With the interaction satisfying the BT constraints, this construction
provides a Poincar\'e-consistent description of the two-body system.
In the LFQM implementation adopted here, the one-body current kernel
is built from on-shell constituent spinors, while the binding dynamics
are encoded in the radial wave functions. This is a three-dimensional
relativistic quantum-mechanical description rather than an exact
reduction of the four-dimensional BS triangle considered here.
The sum of the on-shell constituent LF energies is determined by
the free pair mass $M_0$, whereas the physical meson mass $M$ is
an eigenvalue of the interacting mass operator. These distinct
roles of $M_0$ and $M$ are central to the form-factor projections
discussed below.

Previous work has examined BT-related self-consistency in decay
constants, distribution amplitudes, electromagnetic and scalar form
factors, transverse-momentum-dependent distributions, semileptonic
transitions, and gravitational form factors~\cite{CJ14,CJ15,CJ17,Arifi23a,Arifi23b,Arifi25,CJ24pi,KaonTMD,BDchoi,ChoiSon25}.
The $M\to M_0$ replacement was introduced in Ref.~\cite{CJ14}
as part of a self-consistent correspondence between covariant BS
expressions and the standard LFQM. Subsequent studies applied this
correspondence to additional observables and examined its
self-consistency~\cite{QC18,ChangLiWang19}.
For $P\to P$ transitions, Ref.~\cite{BDchoi} obtained self-consistent
LFQM form factors through a correspondence with the covariant BS
model. The on-shell contributions were first isolated in the BS
amplitudes and then mapped to the standard LFQM by replacing the
LF vertex functions and applying
$M^{(\prime)}\to M_0^{(\prime)}$ throughout the form-factor
integrands. The resulting $(+,\perp)$ and $(+,-)$ current-component
extractions were shown numerically to yield the same $f_-(q^2)$.
Both results were obtained from the mapped on-shell contributions
without separate instantaneous or zero-mode terms.

We call a projection partially BT-based (pBT) when an on-shell BT
current overlap is combined with physical meson masses in the
mass-dependent Lorentz extraction factors. The full BT-based (fBT)
prescription instead uses $M_0$ and $M_0'$ in those internal factors,
point by point under the overlap integral. The physical masses are
retained in the external hadron kinematics and in the definition of
$f_0$. We define the BT completion as the difference between the
fBT and pBT minus-current projections. It is determined once
the wave functions, masses, and pBT baseline are specified, and its
contribution is already included in the fBT result.

In this terminology, the self-consistent LFQM expressions of
Ref.~\cite{BDchoi} already incorporate the on-shell current and
constituent-mass projection structure of the fBT prescription.
The present work develops a direct BT formulation of these
correspondence-based results. Starting from the BT meson states
and the on-shell one-body current, we construct the helicity
kernels and implement the constituent kinematics consistently
in the internal form-factor projections. This makes explicit
the BT structure underlying the earlier correspondence.
For spacelike momentum transfer, we establish current-component independence analytically for
rotationally invariant radial wave functions with the LF Jacobian
and the adopted pseudoscalar spin coupling, provided the reduced
overlaps converge.
The analysis therefore supplies both an analytic foundation
for the earlier numerical agreement and explicit conditions
under which it holds.

The two realizations allow us to identify zero-mode
contributions to $f_-$ in different ways. In the covariant BS
model, we calculate the nonvalence endpoint contribution to the
$(+,\perp)$ extraction directly. Adding it to the valence
contribution reproduces the independently calculated covariant
form factor. The valence-only BT-based LFQM has no explicit
nonvalence sector for the same endpoint calculation. Here
current-component independence provides an internal criterion.
The $(+,\perp)$ extraction is zero-mode free and common to pBT
and fBT. The independently evaluated fBT $(+,-)$ extraction
reproduces this transverse result, whereas the pBT $(+,-)$
extraction generally does not. The BT completion therefore gives the exact integrated correction
needed to restore agreement with the common transverse result.
Within the valence-only BT-based LFQM, this consistency test
indirectly identifies the completion as the effective zero-mode
contribution missing from the pBT minus-current extraction.

The common transverse result also provides a test of whether
the BS prescription for $f_-$ carries over to the BT-based LFQM.
Using the standard-vertex correspondence associated with
Jaus's prescription~\cite{Jaus99}, we retain both the BS valence
current and the effective operator for the nonvalence endpoint
contribution. Even with the same constituent masses and Gaussian
wave functions as in the BT calculation, the resulting $(+,\perp)$
extraction fails to reproduce the component-independent BT result.

The BT construction specifies the meson states, while the
form-factor calculation also requires a choice of current.
Agreement between the transverse and minus extractions tests
the consistency of the adopted one-body current with the fBT
projection. This agreement does not by itself establish full
Lorentz covariance of the truncated current or determine a
unique interacting current.

Section~\ref{sec:kinematics} sets out the kinematics and the
distinction between the BS and BT current realizations.
Section~\ref{sec:bs-benchmark} establishes the exact BS benchmark.
Section~\ref{sec:bt-lfqm} develops the BT extraction, its
effective completion, and the comparison with the standard-vertex
prescription. It also introduces a mass-weighted complex
transverse contour for direct timelike continuation and illustrates
the method with the $D\to K$ form factors.
Section~\ref{sec:summary} summarizes the theoretical implications.
Appendices~\ref{app:covariant-detail}--\ref{app:direct-zm-detail}
contain the BS reductions. Appendix~\ref{app:radial-identity}
summarizes the radial-independent cancellation, while
Appendix~\ref{app:direct-continuation} verifies the regularity
and convergence of the Gaussian overlaps on the timelike contour
and reports the numerical endpoint checks.

\section{Kinematics and the BS--BT distinction}
\label{sec:kinematics}

\subsection{Covariant decomposition of the weak-current matrix element}

We consider a pseudoscalar-to-pseudoscalar semileptonic transition from
an initial meson with momentum $P$ and mass $M$ to a final meson with
momentum $P'$ and mass $M'$. The four-momentum transfer is defined as
$q=P-P'$, and the corresponding weak vector-current matrix element is
given by
\begin{equation}
\label{eq:1}
\Braket{{\cal J}^\mu_W}
\equiv
\langle P'|\bar q_2\gamma^\mu q_1|P\rangle.
\end{equation}

Lorentz covariance allows this matrix element to be decomposed in terms
of the two independent four-vectors
${\cal P}^\mu=(P+P')^\mu$ and $q^\mu=(P-P')^\mu$. One convenient
parametrization is
\begin{equation}
\label{eq:2}
\Braket{{\cal J}^\mu_W}
=
\left[
{\cal P}^\mu
-
q^\mu\frac{M^2-M'^2}{q^2}
\right]F(q^2)
+
q^\mu\frac{M^2-M'^2}{q^2}H(q^2).
\end{equation}
The first Lorentz structure is transverse to $q^\mu$, whereas the second
describes the longitudinal contribution. Indeed, using
$q\cdot{\cal P}=M^2-M'^2$, one finds
$q_\mu\Braket{{\cal J}^\mu_W}=(M^2-M'^2)H(q^2)$.
Thus, $F(q^2)$ and $H(q^2)$ characterize the transverse and
longitudinal parts of the weak-current matrix element, respectively.
The longitudinal contribution is generally nonvanishing because the
flavor-changing vector current is not conserved for unequal
active-quark masses.

The matrix element is more conventionally parametrized in terms of the
weak transition form factors $f_+(q^2)$ and $f_-(q^2)$ as
\begin{equation}
\label{eq:3}
\Braket{{\cal J}^\mu_W}
=
f_+(q^2)\,{\cal P}^\mu
+
f_-(q^2)\,q^\mu.
\end{equation}
Comparing Eqs.~\eqref{eq:2} and \eqref{eq:3}, we obtain
\begin{equation}
\label{eq:4}
f_+(q^2)=F(q^2),
\qquad
f_-(q^2)
=
\frac{M^2-M'^2}{q^2}
\left[H(q^2)-F(q^2)\right].
\end{equation}

It is also useful to introduce the scalar form factor $f_0(q^2)$,
defined by
\begin{equation}
\label{eq:05}
f_0(q^2)
=
f_+(q^2)
+
\frac{q^2}{M^2-M'^2}f_-(q^2).
\end{equation}
Together with Eq.~\eqref{eq:4}, this definition gives
$f_0(q^2)=H(q^2)$, while $f_+(q^2)=F(q^2)$. The parametrizations
$\{F,H\}$, $\{f_+,f_-\}$, and $\{f_+,f_0\}$ are therefore equivalent.

Since the coefficient of $q^\mu$ in Eq.~\eqref{eq:2} is proportional to
$\bigl[H(q^2)-F(q^2)\bigr]/q^2$, regularity of the full matrix element
at $q^2=0$ requires the kinematic constraint $f_0(0)=f_+(0)$,
or equivalently, $H(0)=F(0)$. In the following, we use
$f_+(q^2)$ and $f_-(q^2)$ as the primary transition form factors, with
$f_0(q^2)$ denoting the corresponding scalar combination.

\subsection{Drell--Yan--West frame}

We use LF coordinates $p=(p^+,p^-,{\bm p}_\perp)$, with
$p^\pm=p^0\pm p^3$ and $a\cdot b=\frac{1}{2}\left(a^+b^-+a^-b^+\right) -{\bm a}_\perp\cdot{\bm b}_\perp$.
Working in the Drell--Yan--West (DYW) frame defined by $q^+=0$, we
choose ${\bm P}_\perp={\bm0}_\perp$. The external four-momenta are then
given by
\bea
\label{eq:frame}
P
&=&
\left(
P^+,
\frac{M^2}{P^+},
{\bm0}_\perp
\right),
\qquad
P'
=
\left(
P^+,
\frac{M'^2+{\bm q}_\perp^2}{P^+},
-{\bm q}_\perp
\right),
\nonumber\\
q
&=&
\left(
0,
\frac{M^2-M'^2-{\bm q}_\perp^2}{P^+},
{\bm q}_\perp
\right).
\eea
Defining $Q^2\equiv{\bm q}_\perp^2\geq0$, one finds
$q^2=-Q^2\leq0$. Thus, the DYW frame gives direct access only to the
spacelike form factors $f_\pm(Q^2)$.
Since the integrands depend on ${\bm q}_\perp$ only through rotational
invariants, the integrated form factors depend on $Q^2$ alone, and
${\bm q}_\perp$ may be taken along a fixed transverse axis. The form factors in the physical timelike semileptonic region
are then obtained by analytically continuing the spacelike expressions
according to $Q^2\longrightarrow -q^2$ and $0\leq q^2\leq q^2_{\rm max}=(M-M')^2$.
For the exactly solvable covariant BS model, the validity of this
continuation is verified by the agreement of the analytically
continued LF results with the corresponding manifestly covariant
expressions. For the BT-based LFQM, we evaluate the continued overlap directly at
each requested $q^2$ with the contour introduced in
Sec.~\ref{sec:direct-continuation} and examined in
Appendix~\ref{app:direct-continuation}. The endpoint values are not
obtained from a fit or extrapolation.

\subsection{Current-component extractions}

In the DYW frame, the condition $q^+=0$ allows the form factor
$f_+(q^2)$ to be extracted solely from the plus component of the weak
current:
\begin{equation}
\label{eq:extract-fplus}
f_+(q^2)
=
\frac{\Braket{{\mathcal J}_W^+}}{{\mathcal P}^+}.
\end{equation}
In both models considered here, the plus-current matrix element is
zero-mode free, so its valence contribution suffices to determine
$f_+(q^2)$.

Once $f_+(q^2)$ has been fixed from
Eq.~\eqref{eq:extract-fplus}, the second form factor $f_-(q^2)$ can be
extracted by combining the plus current with either the transverse or
the minus component of the weak current. We denote the two resulting
extractions by $f_-^{(\perp)}(q^2)$ and $f_-^{(-)}(q^2)$,
respectively:
\begin{align}
f_-^{(\perp)}(q^2)
&=
\frac{\Braket{{\mathcal J}_W^+}}{{\mathcal P}^+}
+
\frac{
\langle {\bm{\mathcal J}}_W^\perp \rangle
\cdot {\bm q}_\perp
}{
{\bm q}_\perp^2
},
\label{eq:extract-fperp}
\\
f_-^{(-)}(q^2)
&=
\frac{\Braket{{\mathcal J}_W^-}}{q^-}
-
\frac{{\mathcal P}^-}{q^-}
\frac{\Braket{{\mathcal J}_W^+}}{{\mathcal P}^+}.
\label{eq:extract-fminus}
\end{align}
Here, the superscripts $(\perp)$ and $(-)$ specify the second current
component used together with the plus current. They therefore label
two different extraction procedures for the same physical form factor
and do not represent independent form factors. The mass-dependent
Lorentz projection factors entering the $(+,-)$ extraction are
\begin{equation}
\label{eq:minus-projection-factors}
q^-
=
\frac{M^2-M'^2-{\bm q}_\perp^2}{P^+},
\qquad
{\mathcal P}^-
=
\frac{M^2+M'^2+{\bm q}_\perp^2}{P^+}.
\end{equation}

For a covariant weak-current matrix element, the two extraction
procedures must yield the same physical form factor once all
contributions relevant to the corresponding current components,
including possible LF zero modes, are incorporated consistently:
\begin{equation}
\label{eq:component-independence}
f_-^{(\perp)}(q^2) = f_-^{(-)}(q^2) = f_-(q^2).
\end{equation}

Equation~\eqref{eq:component-independence} is the internal
current-component consistency condition used below. It is a relation
between integrated form factors, not necessarily between their
pointwise kernels. In a one-body-current approximation, satisfying this
condition is necessary for a consistent extraction of $f_-$, but is not
by itself a proof of full Lorentz covariance. The distinction between
an exact algebraic projector identity and a tested overlap relation
will be kept explicit in the BT analysis.

\subsection{Off-shell and on-shell realizations of the same current}

Both the covariant BS model and the BT-based LFQM evaluate the matrix
element of the same one-body weak current,
${\mathcal J}_W^\mu=\bar q_2\gamma^\mu q_1$.
The distinction between the two formulations therefore does not concern
the body number of the current operator. Rather, it lies in the
dynamical description of the bound state and in the realization of the
constituent four-momenta within the one-body current matrix element.

In the four-dimensional covariant BS model, the one-body current
matrix element is represented by a triangle loop involving covariant
meson vertices and constituent propagators that are generally off
their mass shells. With the momentum assignments
$p_1=P-k$ and $p_2=P'-k$, one has $p_1-p_2=q$
for all four components. The physical external momenta and masses
appearing in Eq.~\eqref{eq:minus-projection-factors} are therefore
consistent with the four-dimensional loop kinematics. The difference
between the physical meson masses and the free constituent invariant
masses is accommodated by the off-shell parts of the internal
propagators and by the covariant bound-state vertices.

For $q^+>0$, the LF $k^-$ integration separates the covariant
triangle amplitude into contributions from the valence and nonvalence
regions. The LF decomposition of the fermion propagators also gives
instantaneous terms, including a contribution to the transverse
current in the valence region. In the limit $q^+\to0^+$, a finite part of the
nonvalence contribution may survive as the longitudinal interval
shrinks to an endpoint. This surviving term is the LF zero mode. It is
not an additional many-body current, but a contribution within the LF
representation of the same one-body covariant amplitude required to
reproduce the complete covariant result.

In the BT-based LFQM, by contrast, the same one-body current is
evaluated in the impulse approximation as an overlap of the initial
and final valence LFWFs, with the current kernel constructed from
on-shell constituent spinors. The interaction dependence is assigned
to the BT mass operator and the LFWFs, while no explicit off-shell
propagator, $k^-$ integration, or nonvalence loop appears in the
current matrix element.
With the spectator-antiquark momentum kept unchanged at the weak
vertex, the active-quark momentum transfer satisfies
$p_1^+-p_2^+=q^+=0$ and
${\bm p}_{1\perp}-{\bm p}_{2\perp}={\bm q}_\perp$.
Thus, its plus and transverse components coincide with the
corresponding external momentum-transfer components. The minus
component, however, is fixed by the on-shell constituent momenta and
is given by
\begin{equation}
\label{eq:BT-free-minus-factors}
q_{\rm on}^-
\equiv
p_1^--p_2^-
=
\frac{M_0^2-M_0'^2-{\bm q}_\perp^2}{P^+},
\qquad
{\cal P}_{\rm on}^-
=
\frac{M_0^2+M_0'^2+{\bm q}_\perp^2}{P^+},
\end{equation}
where $M_0$ and $M_0'$ are the free invariant masses of the
initial and final quark--antiquark systems, respectively. In general,
these quantities differ from their physical hadron-level counterparts:
\begin{align}
q^- - q_{\rm on}^-
&=
\frac{
(M^2-M_0^2)-(M'^2-M_0'^2)
}{P^+},
\\
{\mathcal P}^- - {\cal P}_{\rm on}^-
&=
\frac{
(M^2-M_0^2)+(M'^2-M_0'^2)
}{P^+}.
\end{align}
These differences represent the interaction-dependent mass
contributions that are contained in the BT mass operator and the
LFWFs rather than in off-shell constituent propagators.

The distinction is already visible at the constituent spinor level.
Let $j_{21}^{\mu}=\bar u_2(p_2)\gamma^{\mu}u_1(p_1)$ and
$q_{\rm on}=p_1-p_2$, with both active quarks on shell. Then
\begin{equation}
\begin{aligned}
q_{{\rm on}\mu}j_{21}^{\mu}
&=(m_1-m_2)\bar u_2(p_2)u_1(p_1),\\
(q_{\mu}-q_{{\rm on}\mu})j_{21}^{\mu}
&=\frac{q^- -q_{\rm on}^-}{2}\,j_{21}^{+}.
\end{aligned}
\label{eq:constituent-energy-identity}
\end{equation}
The first relation follows from the Dirac equations. The second is
purely kinematic and holds because $q$ and $q_{\rm on}$ have equal plus
and transverse components. Together, these relations show that
replacing $q_{\rm on}$ by the physical momentum transfer $q$ in the
free-spinor divergence identity introduces the additional term
$(q^- -q_{\rm on}^-)j_{21}^{+}/2$.

Because $q^+=q_{\rm on}^+=0$, neither contraction contains
$j_{21}^{-}$. These relations therefore do not determine the minus
current or establish the equality of the transverse- and minus-current
form-factor extractions. For spacelike momentum transfer, Appendix~\ref{app:radial-identity} establishes this equality for the fBT minus-current projection with the specified radial and spin structure.

The fBT prescription uses $q_{\rm on}^-$ and ${\cal P}_{\rm on}^-$ in the
internal projection. The extracted form factors are then combined
with the physical external four-vectors in Eq.~\eqref{eq:3}.
This separates the constituent current kinematics from the binding
contributions encoded in the BT mass operator and radial functions.

\section{Covariant BS model as an exact benchmark}
\label{sec:bs-benchmark}

\subsection{Manifestly covariant calculation}

Within the solvable covariant BS model formulated in
$(3+1)$-dimensional fermion field theory~\cite{BCJ01,BCJ03,MF}, the
one-body weak-current matrix element is represented by a covariant
fermion triangle loop and can be evaluated explicitly.

To regularize the loop integral, the pointlike quark-level weak vertex
is dressed on the initial and final active-quark legs according to~\cite{BCJ01,BCJ03}
\begin{equation}
\Gamma_W^\mu
=
\left(\frac{\Lambda_1^2}{N_{\Lambda_1}}\right)
\gamma^\mu(1-\gamma_5)
\left(\frac{\Lambda_2^2}{N_{\Lambda_2}}\right),
\end{equation}
where
\begin{equation}
N_{\Lambda_1}
=
p_1^2-\Lambda_1^2+i\epsilon,
\qquad
N_{\Lambda_2}
=
p_2^2-\Lambda_2^2+i\epsilon.
\end{equation}
The parameters $\Lambda_1$ and $\Lambda_2$ are regulator masses. After
partial-fraction decomposition~\cite{BCJ01,BCJ03}, the smeared vertex generates a
Pauli--Villars-type subtraction that renders the covariant loop finite.
For a pseudoscalar-to-pseudoscalar transition, the axial-vector current
matrix element vanishes by Lorentz covariance and parity. Consequently,
only the vector-current part contributes, and the regularized matrix
element is given by
\be
\label{eq:31}
\Braket{{\mathcal J}_W^\mu}
=
ig_1g_2\Lambda_1^2\Lambda_2^2
\int\frac{d^4k}{(2\pi)^4}
\frac{S^\mu}
{N_{\Lambda_1}N_1N_{\bar q}N_2N_{\Lambda_2}}.
\ee
Here, $g_1$ and $g_2$ are the meson--quark--antiquark coupling
constants for the initial and final pseudoscalar mesons, respectively.
They are fixed independently by imposing the charge-normalization
condition on the corresponding elastic meson form factors at $q^2=0$.
The propagator denominators are
\begin{equation}
N_1=p_1^2-m_1^2+i\epsilon,
\qquad
N_{\bar q}=k^2-m_{\bar q}^2+i\epsilon,
\qquad
N_2=p_2^2-m_2^2+i\epsilon,
\end{equation}
where $p_1=P-k$ and $p_2=P'-k$.
The masses $m_1$, $m_{\bar q}$, and $m_2$ denote those of the initial
active quark, spectator antiquark, and final active quark,
respectively.

The numerator trace in Eq.~\eqref{eq:31} is
\begin{align}
\label{eq:33}
S^\mu
&=
\operatorname{Tr}
\left[
\gamma_5(\slashed p_1+m_1)
\gamma^\mu
(\slashed p_2+m_2)
\gamma_5(-\slashed k+m_{\bar q})
\right]
\nonumber\\
&=
4(k\cdot P'-k^2+m_2m_{\bar q})P^\mu
+
4(k\cdot P-k^2+m_1m_{\bar q})P'^\mu
\nonumber\\
&\quad
+
4\left(
k^2-P\cdot P'
-m_1m_{\bar q}
-m_2m_{\bar q}
+m_1m_2
\right)k^\mu.
\end{align}

Using Feynman parametrization and comparing the coefficients of
${\mathcal P}^\mu$ and $q^\mu$ with the covariant decomposition in
Eq.~\eqref{eq:3}, we obtain the full covariant form factors
\begin{align}
f_+(q^2)&=\frac{\Npv}{8\pi^2}\int_0^1dx\int_0^{1-x}dy
\Biggl\{
\Bigl[(1-x-y)^2(xM^2+yM'^2)+xy(2-x-y)q^2
\nonumber\\
&\qquad +(x+y)\bigl(m_1m_2-(m_1+m_2)m_{\bar q}\bigr)
+m_{\bar q}(m_1+m_2)\Bigr]\mathcal C
+[3(x+y)-4]\ln\!\left(
\frac{C_{\Lambda_1\Lambda_2}C_{m_1m_2}}
{C_{\Lambda_1m_2}C_{m_1\Lambda_2}}\right)
\Biggr\},\label{eq:36}\\[1ex]
f_-(q^2)&=\frac{\Npv}{8\pi^2}\int_0^1dx\int_0^{1-x}dy
\Biggl\{
\Bigl[(x^2-y^2)(xM^2+yM'^2)+(yM'^2-xM^2)-xy(x-y)q^2
\nonumber\\
&\qquad +(x-y)\bigl(m_1m_2-(m_1+m_2)m_{\bar q}\bigr)
+m_{\bar q}(m_2-m_1)\Bigr]\mathcal C
+3(x-y)\ln\!\left(
\frac{C_{\Lambda_1\Lambda_2}C_{m_1m_2}}
{C_{\Lambda_1m_2}C_{m_1\Lambda_2}}\right)
\Biggr\}.\label{eq:37}
\end{align}
The details of the Feynman-parameter reduction are summarized in
Appendix~\ref{app:covariant-detail}. In Eqs.~\eqref{eq:36} and
\eqref{eq:37}, $x$ and $y$ are Feynman parameters in the triangular domain
$0\leq x\leq1$ and $0\leq y\leq1-x$. The remaining quantities
are defined by
\begin{align}
\label{eq:cov-definitions}
C_{ab}
&=
xa^2+yb^2+(1-x-y)m_{\bar q}^2
-(1-x-y)(xM^2+yM'^2)-xyq^2,
\nonumber\\
{\cal C}
&=
\frac{1}{C_{\Lambda_1\Lambda_2}}
-\frac{1}{C_{\Lambda_1m_2}}
-\frac{1}{C_{m_1\Lambda_2}}
+\frac{1}{C_{m_1m_2}},
\nonumber\\
\Nraw&=g_1g_2\Lambda_1^2\Lambda_2^2,
\nonumber\\
\Npv&=\frac{\Nraw}
{(\Lambda_1^2-m_1^2)(\Lambda_2^2-m_2^2)}.
\end{align}
Here $a\in\{m_1,\Lambda_1\}$ and $b\in\{m_2,\Lambda_2\}$.
The mass-independent ultraviolet-divergent and scheme-dependent
constants cancel in the alternating Pauli--Villars combination.
\subsection{LF valence contributions}

Working in the DYW frame introduced in
Sec.~\ref{sec:kinematics}, we decompose the trace $S^\mu$ in
Eq.~\eqref{eq:33} into an on-shell propagating part,
$S^\mu_{\rm on}$, and an instantaneous part,
$S^\mu_{\rm inst}$. For an internal fermion line carrying
four-momentum $p$, we use the numerator decomposition
\begin{equation}
\slashed{p}+m
=
\left(\slashed{p}_{\rm on}+m\right)
+
\frac{1}{2}\gamma^+
\left(p^- - p_{\rm on}^-\right),
\end{equation}
where $p_{\rm on}$ has the same plus and transverse components as
$p$ and satisfies $p_{\rm on}^2=m^2$. This decomposition is applied
with $p=p_1$ and $p=p_2$ for the active-quark lines and with $p=-k$
for the spectator-antiquark line. Accordingly,
$S^\mu = S^\mu_{\rm on} + S^\mu_{\rm inst}$.
Explicitly, the on-shell part is given by
\begin{equation}
\label{eq:315}
\begin{aligned}
S^\mu_{\rm on}
&=
4\biggl[
p^\mu_{1{\rm on}}
\left(p_{2{\rm on}}\cdot k_{\rm on}\right)
-
k^\mu_{\rm on}
\left(p_{1{\rm on}}\cdot p_{2{\rm on}}\right)
+
p^\mu_{2{\rm on}}
\left(p_{1{\rm on}}\cdot k_{\rm on}\right)
\\
&\qquad
+
m_2m_{\bar q}\,p^\mu_{1{\rm on}}
+
m_1m_{\bar q}\,p^\mu_{2{\rm on}}
+
m_1m_2\,k^\mu_{\rm on}
\biggr],
\end{aligned}
\end{equation}
whereas the instantaneous part reads
\begin{equation}
\label{eq:316}
\begin{aligned}
S^\mu_{\rm inst}
&=
2\delta_1
\biggl[
p^\mu_{2{\rm on}}k^+
-
p_2^+k^\mu_{\rm on}
+
g^{\mu+}
\left(
p_{2{\rm on}}\cdot k_{\rm on}
+
m_2m_{\bar q}
\right)
\biggr]
\\
&\quad
+
2\delta_2
\biggl[
p^\mu_{1{\rm on}}k^+
-
p_1^+k^\mu_{\rm on}
+
g^{\mu+}
\left(
p_{1{\rm on}}\cdot k_{\rm on}
+
m_1m_{\bar q}
\right)
\biggr]
\\
&\quad
+
2\delta_k
\biggl[
p^\mu_{1{\rm on}}p_2^+
+
p_1^+p^\mu_{2{\rm on}}
-
g^{\mu+}
\left(
p_{1{\rm on}}\cdot p_{2{\rm on}}
-
m_1m_2
\right)
\biggr]
+
2g^{\mu+}k^+\delta_1\delta_2.
\end{aligned}
\end{equation}
Here, $\delta_i=p_i^--p_{i{\rm on}}^-$ $(i=1,2)$ and $\delta_k=k^--k_{\rm on}^-$.
In the DYW frame, the corresponding plus-momentum assignments are
$p_1^+=p_2^+=xP^+$ and $k^+=(1-x)P^+$.

Using Eqs.~\eqref{eq:315} and \eqref{eq:316}, we obtain
\begin{equation}
\label{eq:317}
S^+_{\rm on}
=
\frac{4P^+}{1-x}
\left(
{\bm k}_\perp\cdot{\bm k}'_\perp
+
{\cal A}_1{\cal A}_2
\right),
\qquad
S^+_{\rm inst}=0
\end{equation}
for the plus component, whereas the transverse components are given by
\begin{equation}
\label{eq:318}
\begin{aligned}
S^\perp_{\rm on}
&=
-\frac{2{\bm k}_\perp}{x(1-x)}
\biggl[
2{\bm k}_\perp\cdot{\bm k}'_\perp
+
(1-x)
\left(
{\bm q}_\perp^2+m_1^2+m_2^2
\right)
+
2x^2m_{\bar q}^2
\\
&\qquad\qquad
+
2x(1-x)
\left(
m_1m_{\bar q}
+
m_2m_{\bar q}
-
m_1m_2
\right)
\biggr]
-
\frac{2{\bm q}_\perp}{x(1-x)}
\left(
{\bm k}_\perp^2+{\cal A}_1^2
\right),
\\[1ex]
S^\perp_{\rm inst}
&=
-2P^+
\biggl[
\delta_1{\bm k}'_\perp
+
\delta_2{\bm k}_\perp
+
x\delta_k
\left(
2{\bm k}_\perp+{\bm q}_\perp
\right)
\biggr],
\end{aligned}
\end{equation}
where ${\bm k}'_\perp={\bm k}_\perp+(1-x){\bm q}_\perp$ and
${\cal A}_i=(1-x)m_i+xm_{\bar q}$ $(i=1,2)$.

Equations~\eqref{eq:317} and \eqref{eq:318} show that the
transverse-current trace contains instantaneous contributions, whereas
the plus-current trace does not. The result $S^+_{\rm inst}=0$ is
consistent with the zero-mode-free character of the plus-current
matrix element in the present model. However, the absence of an
instantaneous contribution alone does not establish the absence of a
zero mode. This will be confirmed explicitly below by the endpoint
power counting of the nonvalence contribution in the $q^+\to0$ limit.

In the valence region $0<k^+<P^{\prime+}=P^+$, the
spectator-antiquark propagator has a pole at
\begin{equation}
k^-
=
k^-_{\rm on}
-
\frac{i\epsilon}{k^+},
\qquad
k^-_{\rm on}
=
\frac{{\bm k}_\perp^2+m_{\bar q}^2}{k^+}.
\end{equation}
Since $k^+>0$ in this region, the pole lies in the lower half of the
complex $k^-$ plane. Performing the $k^-$ contour integration in
Eq.~\eqref{eq:31}, with $d^4k=\frac12\,dk^+dk^-d^2{\bm k}_\perp$, yields
\bea
\label{eq:319}
\Braket{{\mathcal J}_W^\mu}_{\rm val}
&=&
\frac{\Nraw}{16\pi^3}
\int_0^1\frac{dx}{1-x}
\int d^2{\bm k}_\perp\,
\chi_1(x,{\bm k}_\perp)
\chi_2(x,{\bm k}'_\perp)
S^\mu_{\rm val}.
\eea
Here, $x=p_1^+/P^+=1-k^+/P^+$.
The factor $\Nraw$ is the unreduced loop normalization defined in
Eq.~\eqref{eq:cov-definitions}. The reduced normalization $\Npv$
includes the two Pauli--Villars mass denominators.
The valence trace is defined by
$S^\mu_{\rm val}=(S^\mu_{\rm on}+S^\mu_{\rm inst})|_{k^-=k^-_{\rm on}}$.
At the spectator pole, $\delta_k=0$, whereas the active-quark
off-shell energy differences $\delta_1$ and $\delta_2$ generally
remain nonzero.

The free invariant masses of the initial and final $q\bar q$ systems
are given by
\be
\label{eq:320}
M_0^2
=
\frac{{\bm k}_\perp^2+m_{\bar q}^2}{1-x}
+
\frac{{\bm k}_\perp^2+m_1^2}{x},
\qquad
M_0^{\prime 2}
=
\frac{{\bm k}_\perp^{\prime 2}+m_{\bar q}^2}{1-x}
+
\frac{{\bm k}_\perp^{\prime 2}+m_2^2}{x}.
\ee
We further define
\begin{equation}
M_{\Lambda_1}^2
=
M_0^2(m_1\to\Lambda_1),
\qquad
M_{\Lambda_2}^{\prime 2}
=
M_0^{\prime 2}(m_2\to\Lambda_2).
\end{equation}
The LF vertex functions obtained after the spectator-pole integration
are then
\be
\label{eq:321}
\chi_1(x,{\bm k}_\perp)
=
\frac{1}
{x^2(M^2-M_0^2)(M^2-M_{\Lambda_1}^2)},
\qquad
\chi_2(x,{\bm k}'_\perp)
=
\frac{1}
{x^2(M^{\prime 2}-M_0^{\prime 2})
(M^{\prime 2}-M_{\Lambda_2}^{\prime 2})}.
\ee

Using Eqs.~\eqref{eq:317} and \eqref{eq:318}, the LF valence
contribution to $f_+(q^2)$ and the corresponding $(+,\perp)$
extraction of $f_-(q^2)$ are obtained as
\bea
\label{eq:322}
\left[f_+(q^2)\right]_{\rm val}
&=&
\frac{\Nraw}{16\pi^3}
\int_0^1\frac{dx}{1-x}
\int d^2{\bm k}_\perp\,
\chi_1(x,{\bm k}_\perp)
\chi_2(x,{\bm k}'_\perp)
\frac{S^+_{\rm val}}{{\mathcal P}^+},
\nonumber\\
\left[f_-^{(\perp)}(q^2)\right]_{\rm val}
&=&
\frac{\Nraw}{16\pi^3}
\int_0^1\frac{dx}{1-x}
\int d^2{\bm k}_\perp\,
\chi_1(x,{\bm k}_\perp)
\chi_2(x,{\bm k}'_\perp)
\left(
\frac{S^+_{\rm val}}{{\mathcal P}^+}
+
\frac{
S^\perp_{\rm val}\cdot{\bm q}_\perp
}{
{\bm q}_\perp^2
}
\right),
\eea
where ${\mathcal P}^+=2P^+$, $S^+_{\rm val}=S^+_{\rm on}$, and
\be
\label{eq:323}
\begin{aligned}
S^\perp_{\rm val}
&=
-2{\bm k}_\perp
\biggl[
M^2+M^{\prime 2}+{\bm q}_\perp^2
-
2(m_1-m_{\bar q})(m_2-m_{\bar q})
\biggr]
\\
&\quad
-
2{\bm q}_\perp
\biggl[
(1-x)M^2+xM_0^2-(m_1-m_{\bar q})^2
\biggr].
\end{aligned}
\ee
Consequently,
\bea
\label{eq:324}
\frac{S^+_{\rm val}}{{\mathcal P}^+}
+
\frac{
S^\perp_{\rm val}\cdot{\bm q}_\perp
}{
{\bm q}_\perp^2
}
&=&
2\biggl\{
-(1-x)M^2
+
(m_2-m_{\bar q}){\cal A}_1
-
m_{\bar q}(m_1-m_{\bar q})
\nonumber\\
&&\quad
+
\frac{{\bm k}_\perp\cdot{\bm q}_\perp}{q^2}
\biggl[
M^2+M^{\prime 2}
-
2(m_1-m_{\bar q})(m_2-m_{\bar q})
\biggr]
\biggr\},
\eea
where $q^2=-{\bm q}_\perp^2$ in the DYW frame. Since only the
$(+,\perp)$ extraction is considered in the BS analysis~\cite{Jaus99} below, we
occasionally use the abbreviated notation
$f_-^{\rm val}(q^2)
\equiv
\left[f_-^{(\perp)}(q^2)\right]_{\rm val}$.
The explicit dependence on the physical meson masses $M$ and $M'$ in
$S^\perp_{\rm val}$ arises from the instantaneous terms proportional
to the active-quark off-shell energy differences $\delta_1$ and
$\delta_2$. By contrast, $S^+_{\rm inst}=0$, and hence
$S^+_{\rm val}=S^+_{\rm on}$.

The Feynman-parameter representations of the valence contributions in
Eq.~\eqref{eq:322} are presented in
Appendix~\ref{app:valence-reduction}.
Comparing the valence contributions in Eq.~\eqref{eq:322} with the
manifestly covariant form factors, we find, for the general
unequal-mass transition considered here,
\begin{equation}
\label{eq:bs-valence-summary}
f_+^{\rm cov}(q^2)
=
\left[f_+(q^2)\right]_{\rm val},
\qquad
f_-^{\rm cov}(q^2)
\neq
\left[f_-^{(\perp)}(q^2)\right]_{\rm val}.
\end{equation}
The first equality confirms that the plus-current valence contribution
provides the complete form factor $f_+(q^2)$. By contrast, the second
relation shows that the $(+,\perp)$ valence extraction does not by
itself reproduce the full covariant form factor $f_-(q^2)$. As
demonstrated in the following subsection, the difference is precisely
the LF zero-mode contribution missing from the valence
transverse-current matrix element.

\subsection{Direct zero mode from the nonvalence region}

To identify a possible zero-mode contribution, we first keep
$q^+>0$ and consider the nonvalence region
$P'^+<k^+<P^+$. In this region, the initial active-quark
propagator and the initial smeared vertex generate the poles
\begin{align*}
k^-
&=
P^-
-
p^-_{1{\rm on}}(m_1)
+
\frac{i\epsilon}{p_1^+},
\\
k^-
&=
P^-
-
p^-_{1{\rm on}}(\Lambda_1)
+
\frac{i\epsilon}{p_1^+},
\end{align*}
respectively, where
$p^-_{1{\rm on}}(\mu)
=(\mu^2+{\bm p}_{1\perp}^2)/p_1^+$.
In the frame used here, ${\bm P}_\perp={\bm0}_\perp$, so that
${\bm p}_{1\perp}^2={\bm k}_\perp^2$. Since $p_1^+>0$ in
the nonvalence region, both poles lie in the upper half of the
complex $k^-$ plane.

We define
$P'^+/P^+=1-q^+/P^+\equiv1-\alpha$ and
$x=p_1^+/P^+=1-k^+/P^+$.
The nonvalence region then corresponds to $0<x<\alpha$.
Introducing $x=\alpha z$ with $0<z<1$ maps the shrinking
nonvalence interval onto the fixed domain $0<z<1$, so that the
limit $\alpha\to0^+$ probes the longitudinal endpoint
$x\sim\alpha\to0$.

In the endpoint limit $\alpha\to0^+$ at fixed $z$, the plus
momenta scale as
\begin{equation}
p_1^+
=
\alpha zP^+,
\qquad
p_2^+
=
-\alpha(1-z)P^+,
\qquad
k^+
=
(1-\alpha z)P^+.
\end{equation}
Thus, $p_1^+$ and $p_2^+$ vanish linearly with $\alpha$,
whereas $k^+$ remains finite. At either upper-half-plane pole,
\begin{equation}
p_1^-
=
p^-_{1{\rm on}}(\mu)
=
\frac{\mu^2+{\bm k}_\perp^2}{\alpha zP^+}
=
{\cal O}\left(\frac{1}{\alpha}\right),
\qquad
\mu\in\{m_1,\Lambda_1\}.
\end{equation}
Consequently,
\begin{align}
k^-
&=
P^--p_1^-
=
{\cal O}\left(-\frac{1}{\alpha}\right),
\nonumber\\
p_2^-
&=
P'^--k^-
=
P'^--P^-+p_1^-
=
{\cal O}\left(\frac{1}{\alpha}\right).
\end{align}
Since $x=\alpha z$, these behaviors are equivalently of order
$1/x$. The potentially singular LF-energy factors in the endpoint
power counting are therefore $p_1^-$, $p_2^-$, and $-k^-$.
The endpoint power counting confirms that the nonvalence plus-current
contribution vanishes as $\alpha\to0^+$, consistent with
$f_+^{\rm cov}=[f_+]_{\rm val}$ in
Eq.~\eqref{eq:bs-valence-summary}. Hence,
$\left[f_+(q^2)\right]_{\rm Z.M.}=0$.

By contrast, the transverse trace contains LF-energy factors that
scale as $1/x$ in the endpoint region. In the power counting of the
complete nonvalence integrand, only terms of order $1/x$ in the
transverse trace can compensate for the shrinking longitudinal
interval. Terms in which a singular LF energy is accompanied by a
vanishing plus-momentum factor are at most of order unity. For example,
$p_{1{\rm on}}^-p_2^+\sim \mathcal{O}(1)$,
and similarly for $p_1^+p_{2{\rm on}}^-$. Such terms therefore do not
generate a zero-mode contribution in the limit $\alpha\to0^+$.

We thus retain only the leading $1/x$ terms in the transverse trace.
From the on-shell part,
\begin{equation}
S^\perp_{\rm on}
=
4\left[
{\bm p}_{1\perp}
(p_{2{\rm on}}\cdot k_{\rm on})
-
{\bm k}_\perp
(p_{1{\rm on}}\cdot p_{2{\rm on}})
+
{\bm p}_{2\perp}
(p_{1{\rm on}}\cdot k_{\rm on})
+\cdots
\right],
\end{equation}
the singular contributions arise from
$p_{2{\rm on}}\cdot k_{\rm on}\supset
p_{2{\rm on}}^-k^+/2$ and
$p_{1{\rm on}}\cdot k_{\rm on}\supset
p_{1{\rm on}}^-k^+/2$.
Since $k^+$ remains finite in the endpoint limit, the leading
on-shell contribution is
\begin{equation}
\left.S^\perp_{\rm on}\right|_{1/x}
=
2k^+
\left(
{\bm p}_{1\perp}p^-_{2{\rm on}}
+
{\bm p}_{2\perp}p^-_{1{\rm on}}
\right).
\end{equation}

The leading instantaneous contribution is
\begin{equation}
\left.S^\perp_{\rm inst}\right|_{1/x}
=
2k^+
\left(
{\bm p}_{2\perp}\delta_1
+
{\bm p}_{1\perp}\delta_2
\right).
\end{equation}
The remaining instantaneous terms are accompanied by the vanishing
plus momenta $p_1^+$ or $p_2^+$ and are therefore at most of order
${\cal O}(1)$ in the endpoint limit.
Combining the leading on-shell and instantaneous contributions, we
obtain
\begin{align}
\left.S^\perp_{\rm nv}\right|_{1/x}
&=
2k^+
\biggl[
{\bm p}_{1\perp}
\left(
p^-_{2{\rm on}}+\delta_2
\right)
+
{\bm p}_{2\perp}
\left(
p^-_{1{\rm on}}+\delta_1
\right)
\biggr]
\nonumber\\
&=
2k^+
\left(
{\bm p}_{1\perp}p_2^-
+
{\bm p}_{2\perp}p_1^-
\right).
\label{eq:singular-transverse-trace}
\end{align}
In the endpoint region,
$p_2^-=p_1^-+{\cal O}(1)$ and $k^+\to P^+$ as $x\to0$.
Using also
${\bm p}_{2\perp}={\bm p}_{1\perp}-{\bm q}_\perp$, we obtain
\begin{equation}
\label{eq:330}
\begin{aligned}
S^\perp_{\rm Z.M.}
&\equiv
\left.S^\perp_{\rm nv}\right|_{1/x}
=
2P^+p_1^-
\left(
2{\bm p}_{1\perp}-{\bm q}_\perp
\right).
\end{aligned}
\end{equation}
The zero-mode numerator therefore has a common tensor structure for
all four terms generated by the Pauli--Villars decomposition in
Eq.~\eqref{eq:34}. The dependence on the mass parameters enters only
when $p_1^-$ and the remaining denominators are evaluated at the pole
appropriate to each term.

To evaluate the nonvalence contribution, we apply the decomposition
in Eq.~\eqref{eq:34} before performing the $k^-$ contour integration.
The amplitude is thereby expressed as four three-denominator terms of
the form $1/(N_{\bar q}N_aN_b)$, with
$a\in\{m_1,\Lambda_1\}$ and
$b\in\{m_2,\Lambda_2\}$, together with the corresponding
Pauli--Villars signs. Each term contains only
one initial active-line denominator $N_a$ and hence a single
upper-half-plane pole, at which $p_1^- =p^-_{1{\rm on}}(a)$.
This supplies the common zero-mode spin structure,
whereas the specific dependence on $m_1$, $\Lambda_1$, $m_2$, and
$\Lambda_2$ arises through the pole value and the remaining
denominators in each Pauli--Villars term.

Substituting the zero-mode numerator in Eq.~\eqref{eq:330} into the
nonvalence amplitude, the zero-mode contribution to the
$(+,\perp)$ extraction of $f_-(q^2)$ is given by
\begin{align}
f_-^{\rm Z.M.}(q^2)
&\equiv
\left[f_-^{(\perp)}(q^2)\right]_{\rm Z.M.}
\nonumber\\
&=
i\Npv
\lim_{\alpha\to0^+}
\int_{\rm nv}
\frac{d^4k}{(2\pi)^4}\,
\frac{
S^\perp_{\rm Z.M.}\cdot{\bm q}_\perp
}{
{\bm q}_\perp^2
}
\nonumber\\
&\quad\times
\biggl[
\frac{1}{N_{\bar q}N_1N_2}
-
\frac{1}{N_{\bar q}N_1N_{\Lambda_2}}
-
\frac{1}{N_{\bar q}N_{\Lambda_1}N_2}
+
\frac{1}{N_{\bar q}N_{\Lambda_1}N_{\Lambda_2}}
\biggr].
\label{eq:331}
\end{align}

The finite zero-mode contribution is
\begin{equation}
\label{eq:fmzm0}
f_-^{\rm Z.M.}(q^2)
=
\frac{\Npv}{8\pi^2}
\int_0^1dz\,(2z-1)
\ln\left(
\frac{
B_{\Lambda_1m_2}B_{m_1\Lambda_2}
}{
B_{\Lambda_1\Lambda_2}B_{m_1m_2}
}
\right),
\end{equation}
where $B_{ab}=z(1-z){\bm q}_\perp^2+(1-z)a^2+zb^2$. The details of
the contour integration and transverse dimensional regularization are
summarized in Appendix~\ref{app:direct-zm-detail}.

Combining the valence and zero-mode contributions, the complete LF
representation of the covariant form factors is
\begin{equation}
\label{eq:bs-full-lf}
f_+^{\rm cov}(q^2)
=
\left[f_+(q^2)\right]_{\rm val},
\qquad
f_-^{\rm cov}(q^2)
=
\left[f_-^{(\perp)}(q^2)\right]_{\rm val}
+
f_-^{\rm Z.M.}(q^2).
\end{equation}
Thus, the plus-current form factor is completely saturated by its
valence contribution, whereas the transverse-current extraction of
$f_-(q^2)$ requires the finite zero-mode contribution generated by
the longitudinal endpoint.

\subsection{Effective zero-mode operator and equivalence with the direct result}

We now construct a valence-space representation of the endpoint
contribution in Eq.~\eqref{eq:fmzm0} using the $\omega$-dependent
covariant decomposition developed by Jaus~\cite{Jaus99}, based on the
covariant formulation of LF dynamics in Refs.~\cite{CDKM,SCC}. The LF orientation is specified by a lightlike
four-vector $\omega^\mu$. With our convention,
$\omega^\mu=(0,2,{\bm0}_\perp)$ gives $\omega\cdot x=x^+=0$.
After the spectator-pole reduction, structures proportional to
$\omega^\mu$ are spurious and must cancel from the physical form
factors. The relevant $C$ coefficients are zero-mode sensitive, while
the residual $B$ coefficients are constrained separately by the
integrated covariance identities.

The required rank-one decomposition is understood under the LF
valence integral rather than as a pointwise identity between the
integrands:
\begin{equation}
\label{eq:j1}
p_1^\mu
\doteq
{\cal P}^\mu A^{(1)}_1
+
q^\mu A^{(1)}_2
+
\frac{\omega^\mu}{\omega\cdot{\cal P}}C^{(1)}_1,
\end{equation}
where the symbol $\doteq$ denotes equality after integration with the
scalar LF vertex functions. In the $q^+=0$ frame, the coefficients are
given by
\begin{align}
A^{(1)}_1
&=
\frac{\omega\cdot p_1}{\omega\cdot{\cal P}}
=
\frac{x}{2},
\nonumber\\
A^{(1)}_2
&=
\frac{1}{q^2}
\left[
p_1\cdot q
-
(q\cdot{\cal P})
\frac{\omega\cdot p_1}{\omega\cdot{\cal P}}
\right]
=
\frac{x}{2}
+
\frac{{\bm k}_\perp\cdot{\bm q}_\perp}{q^2},
\nonumber\\
C^{(1)}_1
&=
p_1\cdot{\cal P}
-
{\cal P}^2A^{(1)}_1
-
(q\cdot{\cal P})A^{(1)}_2
=
Z_2-N_{\bar q},
\label{eq:j2}
\end{align}
where $N_{\bar q}=k^2-m_{\bar q}^2$ and
\begin{equation}
\label{eq:j3}
Z_2
=
x(M^2-M_0^2)
+
m_1^2-m_{\bar q}^2
+
(1-2x)M^2
-
\left(q^2+q\cdot{\cal P}\right)
\frac{{\bm k}_\perp\cdot{\bm q}_\perp}{q^2}.
\end{equation}

Among the coefficients in Eq.~\eqref{eq:j2}, only
$C^{(1)}_1$ depends on the off-shell LF energy $p^-_1$,
through the scalar product $p_1\cdot{\cal P}$. The coefficient
$A^{(1)}_1$ depends only on the plus momentum, while
$A^{(1)}_2$ is independent of $p^-_1$ because the term
proportional to $p^-_1q^+$ in $p_1\cdot q$ vanishes in the
$q^+=0$ frame. Since ${\cal P}^+=2P^+$, the part of
$p_1\cdot{\cal P}$ proportional to the off-shell LF energy is
$P^+p^-_1$.

The term proportional to $C^{(1)}_1$ multiplies the
LF-orientation vector $\omega^\mu$ and therefore represents a
spurious orientation-dependent contribution. Jaus's original
prescription concerns the spectator off-shellness
$N_{\bar q}$, which corresponds to $\widehat N_2$ in the notation
of Ref.~\cite{Jaus99}. In our notation, the integrated form of
his zero-mode prescription is
\begin{equation}
\label{eq:j5}
I[N_{\bar q}]_{\rm Z.M.}
=
I[Z_2]_{\rm val}
=
\frac{\Nraw}{16\pi^3}
\int_0^1\frac{dx}{1-x}
\int d^2{\bm k}_\perp\,
\chi_1(x,{\bm k}_\perp)
\chi_2(x,{\bm k}'_\perp)
Z_2.
\end{equation}
At the spectator pole, $N_{\bar q}=0$, so that
$I[N_{\bar q}]_{\rm val}=0$ and
$I[N_{\bar q}]_{\rm full}=I[N_{\bar q}]_{\rm Z.M.}$.
Thus Eq.~\eqref{eq:j5} identifies a zero-mode
contribution rather than an ordinary spectator-pole term.
This relation is implemented through Jaus's effective replacement
\begin{equation}
\label{eq:Jaus-N-to-Z}
N_{\bar q}
\longrightarrow
Z_2
\end{equation}
in the zero-mode-sensitive part of the valence integrand. Since
$C^{(1)}_1=Z_2-N_{\bar q}$ multiplies the spurious
$\omega^\mu$-dependent structure, Eqs.~\eqref{eq:j5} and
\eqref{eq:Jaus-N-to-Z} remove the contribution of
$C^{(1)}_1$ from the zero-mode-completed amplitude. The replacement
in Eq.~\eqref{eq:Jaus-N-to-Z} is Jaus's original zero-mode
prescription.

The direct nonvalence calculation performed in the present work
provides an interpretation of Jaus's prescription in terms of the
off-shell LF energy factor $P^+p^-_1$. In the endpoint region, the
singular part of the spectator off-shellness behaves as
$N_{\bar q}|_{\rm sing.}\simeq-P^+p^-_1$. The corresponding
endpoint calculation gives
\begin{equation}
\label{eq:pminus-N-relation}
I[P^+p^-_1]_{\rm Z.M.}
=
-I[N_{\bar q}]_{\rm Z.M.}.
\end{equation}
This identity is a result of our direct nonvalence calculation and
is not an additional prescription assumed from the outset.

Combining Eq.~\eqref{eq:pminus-N-relation} with Jaus's relation
in Eq.~\eqref{eq:j5} gives
$I[P^+p^-_1]_{\rm Z.M.}=-I[Z_2]_{\rm val}$. Therefore, within
the zero-mode-sensitive valence convolution of the present solvable
BS model, Jaus's prescription
$N_{\bar q}\to Z_2$ can be interpreted in terms of the off-shell
LF energy as
\begin{equation}
\label{eq:pminus-to-Z}
P^+p^-_1
\longrightarrow
-Z_2.
\end{equation}
Equation~\eqref{eq:pminus-to-Z} is thus not an independent
prescription proposed by Jaus. It is our effective representation of
his original prescription, obtained by combining
$N_{\bar q}\to Z_2$ with the direct zero-mode identity in
Eq.~\eqref{eq:pminus-N-relation}.

The term proportional to $P^+p^-_1{\bm q}_\perp$ in
Eq.~\eqref{eq:330} can therefore be incorporated directly into the
valence convolution using Eq.~\eqref{eq:pminus-to-Z}. The term
proportional to $P^+p^-_1{\bm p}_{1\perp}$, however, involves a
rank-two tensor structure and requires the corresponding covariant
tensor decomposition.

For this purpose, we use the tensor decomposition introduced by
Jaus~\cite{Jaus99},
\begin{align}
\label{eq:j6}
p_1^\mu p_1^\nu
&\doteq
g^{\mu\nu}A^{(2)}_1
+
{\cal P}^\mu{\cal P}^\nu A^{(2)}_2
+
\left(
{\cal P}^\mu q^\nu
+
q^\mu{\cal P}^\nu
\right)A^{(2)}_3
+
q^\mu q^\nu A^{(2)}_4
\nonumber\\
&\quad
+
\frac{1}{\omega\cdot{\cal P}}
\left(
{\cal P}^\mu\omega^\nu
+
\omega^\mu{\cal P}^\nu
\right)B^{(2)}_1
\nonumber\\
&\quad
+
\frac{1}{\omega\cdot{\cal P}}
\left(
q^\mu\omega^\nu
+
\omega^\mu q^\nu
\right)C^{(2)}_1
+
\frac{\omega^\mu\omega^\nu}
{(\omega\cdot{\cal P})^2}
C^{(2)}_2,
\end{align}
where, as in Eq.~\eqref{eq:j1}, the symbol $\doteq$ denotes equality
under the LF valence integral. The coefficients are
\begin{align}
\label{eq:j7}
A^{(2)}_1
&=
-{\bm k}_\perp^2
-
\frac{
\left(
{\bm k}_\perp\cdot{\bm q}_\perp
\right)^2
}{q^2},
\qquad
A^{(2)}_2
=
\left[A^{(1)}_1\right]^2,
\nonumber\\
A^{(2)}_3
&=
A^{(1)}_1A^{(1)}_2,
\qquad
A^{(2)}_4
=
\left[A^{(1)}_2\right]^2
-
\frac{1}{q^2}A^{(2)}_1,
\nonumber\\
B^{(2)}_1
&=
A^{(1)}_1C^{(1)}_1
-
A^{(2)}_1,
\nonumber\\
C^{(2)}_1
&=
A^{(1)}_2C^{(1)}_1
+
\frac{q\cdot{\cal P}}{q^2}A^{(2)}_1,
\nonumber\\
C^{(2)}_2
&=
\left[C^{(1)}_1\right]^2
+
\left[
{\cal P}^2
-
\frac{(q\cdot{\cal P})^2}{q^2}
\right]A^{(2)}_1.
\end{align}

The coefficients $A_i^{(2)}$ multiply the physical,
$\omega$-independent tensor structures. We note that $B_1^{(2)}$, although $\omega$ dependent, is not
zero-mode sensitive and is treated separately through Jaus's
integrated covariance identity~\cite{Jaus99}.
The coefficient $C_1^{(2)}$, by contrast, contains the
zero-mode-sensitive rank-two structure required for the present
transverse-current matrix element. Jaus incorporates its zero-mode
contribution by imposing $C_1^{(2)}\to0$, which leads to his
rank-two prescription
\begin{equation}
\label{eq:Jaus-rank2}
A^{(1)}_2N_{\bar q}
\longrightarrow
A^{(1)}_2Z_2
+
\frac{q\cdot{\cal P}}{q^2}A^{(2)}_1.
\end{equation}
The rank-one and rank-two replacements are understood under the
covariant-vertex convolution together with the integrated covariance
identities. The coefficient $C_2^{(2)}$ is of second order in the
off-shell LF energy and is not needed for the transverse-current zero
mode considered here.

Combining Jaus's rank-two prescription in
Eq.~\eqref{eq:Jaus-rank2} with the direct zero-mode identity in
Eq.~\eqref{eq:pminus-N-relation}, we find that the tensor zero-mode
contribution can be represented within the valence convolution as
\begin{equation}
\label{eq:pminus-tensor-replacement}
P^+p_1^-{\bm p}_{1\perp}
\longrightarrow
-{\bm q}_\perp
\left[
A^{(1)}_2Z_2
+
\frac{q\cdot{\cal P}}{q^2}A^{(2)}_1
\right].
\end{equation}
Equation~\eqref{eq:pminus-tensor-replacement} is therefore not an
independent prescription proposed by Jaus. It follows from combining
his rank-two replacement in Eq.~\eqref{eq:Jaus-rank2} with the direct
nonvalence zero-mode identity derived in the present BS-model
calculation.

Applying Eqs.~\eqref{eq:pminus-to-Z} and
\eqref{eq:pminus-tensor-replacement} to the direct zero-mode numerator
in Eq.~\eqref{eq:330}, we obtain the following effective
valence-space representation:
\begin{equation}
\label{eq:SZMop}
S^\perp_{\rm Z.M.}
\longrightarrow
S^\perp_{{\rm Z.M.},{\rm eff}}
=
2{\bm q}_\perp
\left\{
Z_2
-
2\left[
A^{(1)}_2Z_2
+
\frac{q\cdot{\cal P}}{q^2}A^{(2)}_1
\right]
\right\}.
\end{equation}
We call $S^\perp_{{\rm Z.M.},{\rm eff}}$ the effective zero-mode
operator for this BS model. The arrow denotes an effective replacement under the LF valence
convolution, rather than a pointwise equality between the direct
zero-mode numerator and the effective operator.

The effective valence-space representation of the transverse-current
zero-mode contribution to $f_-(q^2)$ is therefore
\begin{align}
\label{eq:fmzm}
f_-^{\rm Z.M.}(q^2)
&\equiv
\left[f_-^{(\perp)}(q^2)\right]_{\rm Z.M.}
\nonumber\\
&=
\frac{\Nraw}{16\pi^3}
\int_0^1\frac{dx}{1-x}
\int d^2{\bm k}_\perp\,
\chi_1(x,{\bm k}_\perp)
\chi_2(x,{\bm k}'_\perp)
\frac{
S^\perp_{{\rm Z.M.},{\rm eff}}
\cdot{\bm q}_\perp
}{
{\bm q}_\perp^2
}.
\end{align}
Combining this contribution with the LF valence result, the complete
$(+,\perp)$ extraction in the covariant BS model is
\begin{align}
\label{eq:fmconv}
f_-^{\rm cov}(q^2)
&=
\left[f_-^{(\perp)}(q^2)\right]_{\rm full}
\nonumber\\
&=
\frac{\Nraw}{16\pi^3}
\int_0^1\frac{dx}{1-x}
\int d^2{\bm k}_\perp\,
\chi_1(x,{\bm k}_\perp)
\chi_2(x,{\bm k}'_\perp)
\left[
\frac{S^+_{\rm val}}{{\cal P}^+}
+
\frac{
\left(
S^\perp_{\rm val}
+
S^\perp_{{\rm Z.M.},{\rm eff}}
\right)
\cdot{\bm q}_\perp
}{
{\bm q}_\perp^2
}
\right],
\end{align}
where ${\cal P}^+=2P^+$. The complete spin factor entering this
expression is
\begin{align}
\label{eq:fmspinfull}
&
\frac{S^+_{\rm val}}{{\cal P}^+}
+
\frac{
S^\perp_{\rm val}\cdot{\bm q}_\perp
}{
{\bm q}_\perp^2
}
+
\frac{
S^\perp_{{\rm Z.M.},{\rm eff}}
\cdot{\bm q}_\perp
}{
{\bm q}_\perp^2
}
\nonumber\\
&=
2\biggl\{
-x(1-x)M^2
-{\bm k}_\perp^2
-m_1m_{\bar q}
+(m_2-m_{\bar q}){\cal A}_1
+
2\frac{q\cdot{\cal P}}{q^2}
\left[
{\bm k}_\perp^2
+
2\frac{
({\bm k}_\perp\cdot{\bm q}_\perp)^2
}{q^2}
\right]
+
2\frac{
({\bm k}_\perp\cdot{\bm q}_\perp)^2
}{q^2}
\nonumber\\
&\qquad
+
\frac{
{\bm k}_\perp\cdot{\bm q}_\perp
}{q^2}
\biggl[
M'^2
-(1-x)
\left(
q^2+q\cdot{\cal P}
\right)
+2xM_0^2
-(1-2x)M^2
-2(m_1-m_{\bar q})(m_1+m_2)
\biggr]
\biggr\}.
\end{align}
Equation~\eqref{eq:fmspinfull} follows directly from
Eqs.~\eqref{eq:324} and \eqref{eq:SZMop}.

Equation~\eqref{eq:fmzm0} gives the BS zero mode directly
from the $q^+\to0^+$ limit of the nonvalence region.
Equation~\eqref{eq:fmzm} gives its valence-space representation,
obtained by combining the endpoint identity in
Eq.~\eqref{eq:pminus-N-relation} with Jaus's replacement
$N_{\bar q}\to Z_2$ and its rank-two counterpart in
Eq.~\eqref{eq:Jaus-rank2}. Both replacements are understood
under the valence integral. The two evaluations of the BS endpoint
contribution agree within numerical precision. No corresponding
nonvalence endpoint calculation is available in the valence-only
BT-based LFQM. As we show in Sec.~\ref{sec:bt-lfqm},
current-component independence instead identifies the difference
between the fBT and pBT minus-current projections as the effective
zero-mode correction missing from the pBT extraction.

\subsection{Numerical benchmark in the BS model}

To provide an explicit benchmark for the general unequal-mass case,
in which $m_1$, $m_2$, and $m_{\bar q}$ are all different, we consider
the semileptonic decay $D\to K\ell\nu$. The model parameters used for
the $D$ and $K$ mesons are
$M_D=1.869~{\rm GeV}$, $M_K=0.494~{\rm GeV}$,
$m_c=1.6~{\rm GeV}$, $m_{u(d)}=0.3~{\rm GeV}$,
$m_s=0.5~{\rm GeV}$, $\Lambda_D=1.8~{\rm GeV}$, and
$\Lambda_K=0.9~{\rm GeV}$. The meson--quark coupling constants
$g_D$ and $g_K$ are fixed independently by imposing the
charge-normalization condition on the corresponding elastic meson
form factors at $q^2=0$.

\begin{figure}
\centering
\includegraphics[width=2.8in,height=2.3in]{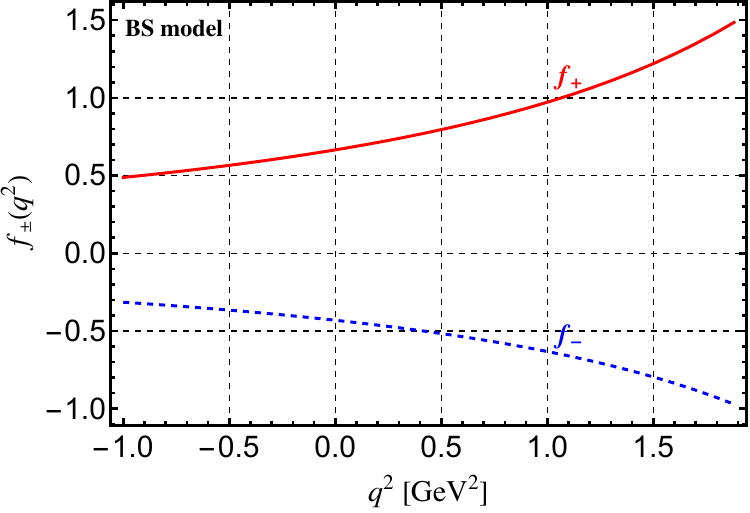}\hspace{1cm}%
\includegraphics[width=2.8in,height=2.3in]{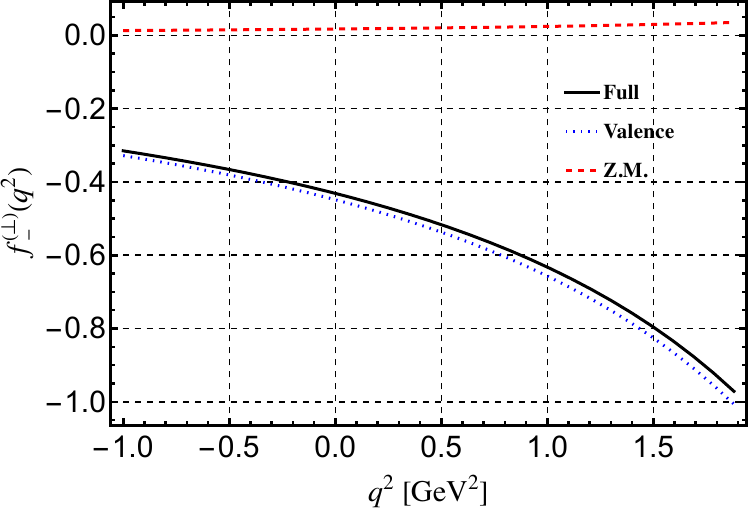}
\caption{The weak form factors $f_{\pm}(q^2)$
(left) and the decomposition of $f_-(q^2)=f_-^{(\perp)}(q^2)$ (right) for the
$D\to K$ semileptonic decay in the exactly solvable covariant BS model.}
\label{fig1}
\end{figure}

Figure~\ref{fig1} shows the weak form factors for the
$D\to K\ell\nu$ decay obtained from the direct covariant and LF
calculations. Although the physical decay region is
$0\leq q^2\leq q_{\rm max}^2=(M_D-M_K)^2$, we extend the displayed
range to $-1~{\rm GeV}^2\leq q^2\leq q_{\rm max}^2$ to show the
continuation from the spacelike region to the physical timelike region.
The left panel shows the full results for $f_+(q^2)$ and
$f_-(q^2)=f_-^{(\perp)}(q^2)$. The right panel decomposes
$f_-^{(\perp)}(q^2)$ into its LF valence and zero-mode contributions.

For $f_+(q^2)$, the manifestly covariant result in
Eq.~\eqref{eq:36} is reproduced by the LF valence contribution,
$f_+^{\rm cov}(q^2)=[f_+(q^2)]_{\rm val}$. For $f_-(q^2)$, the
manifestly covariant result in Eq.~\eqref{eq:37} is reproduced only
after the zero-mode contribution is included,
$f_-^{\rm cov}(q^2)=[f_-^{(\perp)}(q^2)]_{\rm val}
+f_-^{\rm Z.M.}(q^2)$. The magnitude of the zero-mode contribution
increases toward zero recoil and decreases toward maximum recoil.
Its inclusion is required to reproduce the full covariant form factor
$f_-(q^2)$.

The BS analysis thus provides an explicit benchmark for the
nonvalence endpoint contribution and its effective valence-space
representation. In the following section, we turn to the BT-based
LFQM, where the current matrix element is instead constructed from
on-shell constituent spinors. We derive the fBT projection directly
from this constituent kinematics and examine current-component
independence within that realization.

\section{Bakamjian--Thomas construction of the light-front form factors}
\label{sec:bt-lfqm}

\subsection{Meson states and wave functions}

The BT construction is not tied to a particular form of relativistic
dynamics. The same class of $P\to P$ weak transitions has also been
studied within a point-form BT constituent-quark
model~\cite{Heger21}. We test the one-body current using the
component relation in Eq.~\eqref{eq:component-independence}.

In the conventional LFQM
\cite{Jaus90,Jaus91,CCH97,CJ99a,CJ99b,Choi07b,CJLR15,ACJO22},
the meson state is approximated by its lowest Fock component, namely
the constituent quark--antiquark sector. A meson carrying total
four-momentum $P$ is then represented as a relativistic two-body
system composed of a quark and an antiquark. Fock components
containing additional partons are omitted from the explicit state
expansion.

In the BT construction, the interaction dependence is assigned to
the invariant mass operator,
$\hat M=\hat M_0+\hat V_{q\bar q}$, where $\hat M_0$ is the free mass
operator and $\hat V_{q\bar q}$ is an effective interaction between
the constituent quark and antiquark. The latter parametrizes the
nonperturbative dynamics of QCD and typically incorporates both
confining and spin-dependent components
\cite{CJ99a,CJ99b,Choi07b,CJLR15,ACJO22}. When the interaction satisfies
the BT requirements, including invariance under the kinematic
generators and commutation with the noninteracting internal spin, the
mass-operator construction preserves the Poincar\'e algebra
\cite{BT53,Poly10,CCKP}.

The constituent momenta entering the internal wave function are taken
to be on their respective mass shells, while the interaction is
encoded in the LFWF
$\Psi^{JJ_z}_{\lambda_q\lambda_{\bar q}}(x,{\bm k}_\perp)$.
This wave function is the LF momentum representation of the
bound-state eigenvector $\ket{\Psi^{JJ_z}}$, defined by
$\hat M\ket{\Psi^{JJ_z}}=M\ket{\Psi^{JJ_z}}$, where $M$ is the
physical meson mass. The Gaussian radial wave function adopted below
provides a phenomenological realization of this two-body bound state
and is not obtained by solving an explicit four-dimensional BS kernel.

In the valence approximation, a meson with momentum $P$, total spin
$J$, and spin projection $J_z$ is expanded in the two-particle
quark--antiquark basis as
\begin{align}\label{eq:18}
\ket{P;JJ_z}
={}&
\sum_{\lambda_q,\lambda_{\bar q}}
\int [\mathrm d^3{\bm p}_q]\,
     [\mathrm d^3{\bm p}_{\bar q}]\,
2(2\pi)^3
\delta^{(3)}
\bigl({\bm P}-{\bm p}_q-{\bm p}_{\bar q}\bigr)
\nonumber\\
&\times
\Psi^{JJ_z}_{\lambda_q\lambda_{\bar q}}
(x,{\bm k}_\perp)
\ket{q(p_q,\lambda_q)
\bar q(p_{\bar q},\lambda_{\bar q})}.
\end{align}
Here, $p_q$ and $p_{\bar q}$ are the on-mass-shell constituent
four-momenta, while $\lambda_q$ and $\lambda_{\bar q}$ denote their LF
helicities. We use bold symbols such as
${\bm P}\equiv(P^+,{\bm P}_\perp)$ and
${\bm p}\equiv(p^+,{\bm p}_\perp)$ to denote LF three-momenta.
Accordingly,
$\delta^{(3)}({\bm P}-{\bm p}_q-{\bm p}_{\bar q})
=
\delta(P^+-p_q^+-p_{\bar q}^+)\,
\delta^{(2)}({\bm P}_\perp-{\bm p}_{q\perp}
-{\bm p}_{\bar q\perp})$
and the one-particle LF phase-space element is defined by
$[\mathrm d^3{\bm p}]=\frac{\mathrm dp^+\,\mathrm d^2{\bm p}_\perp}{2(2\pi)^3}$.

It is convenient to separate the overall motion of the meson from the
relative motion of its constituents. For each constituent $i$, we
introduce $x_i=\frac{p_i^+}{P^+}$ and
${\bm k}_{i\perp}={\bm p}_{i\perp}-x_i{\bm P}_\perp$,
which obey $\sum_i x_i=1$ and $\sum_i{\bm k}_{i\perp}=0$.
For the quark, we use the convention
$x_q\equiv x$ and ${\bm k}_{q\perp}\equiv-{\bm k}_\perp$.
The overall meson motion is thereby separated from the internal
motion. The LFWF depends only on the boost-invariant internal
variables $x$ and ${\bm k}_\perp$, while the total momentum $P$
enters through the constituent momenta in the Fock basis states.

The free invariant mass $M_0$ of the noninteracting $q\bar q$
pair, defined in Eq.~\eqref{eq:320}, depends only on the internal
variables $x$ and ${\bm k}_\perp$ and determines the free
constituent kinematics entering the BT construction.

The normalization of the physical meson state is fixed by
\begin{equation}\label{eq:statenorm}
\Braket{P'|P} =2(2\pi)^3P^+\,\delta^{(3)}({\bm P}'-{\bm P}).
\end{equation}
For a pseudoscalar meson with \(J=J_z=0\), this convention implies
\begin{equation}
\label{eq:Psinorm}
\sum_{\lambda_q,\lambda_{\bar q}}
\int_0^1 \mathrm dx
\int\frac{\mathrm d^2{\bm k}_\perp}{2(2\pi)^3}
\left|
\Psi^{00}_{\lambda_q\lambda_{\bar q}}
(x,{\bm k}_\perp)
\right|^2
=1.
\end{equation}

We separate the momentum distribution from the helicity structure
of the LFWF by writing
\begin{equation}
\label{eq:5}
\Psi^{JJ_z}_{\lambda_q\lambda_{\bar q}}
(x,{\bm k}_\perp)
=
\phi(x,{\bm k}_\perp)\,
\mathcal{R}^{JJ_z}_{\lambda_q\lambda_{\bar q}}
(x,{\bm k}_\perp).
\end{equation}
The radial wave function $\phi$ describes the internal momentum
distribution, while
$\mathcal{R}^{JJ_z}_{\lambda_q\lambda_{\bar q}}$ describes the
spin--orbit structure in the LF helicity basis. The latter may be obtained from the corresponding
instant-form spin wave function through the interaction-independent
Melosh rotation~\cite{Melosh1974}, which changes the spin basis while
preserving the total angular momentum quantum numbers.

For a $J^P=0^-$ meson, the spin--orbit wave function can be written in the
covariant form~\cite{Jaus90,Jaus91}
\begin{equation}
\label{eq8}
\mathcal{R}^{00}_{\lambda_q\lambda_{\bar q}}
=
-\frac{
\bar u_{\lambda_q}(p_q)\gamma_5
v_{\lambda_{\bar q}}(p_{\bar q})
}{
\sqrt{2}\,
\sqrt{M_0^2-(m_q-m_{\bar q})^2}
}.
\end{equation}
Its helicity sum is normalized to unity:
\begin{equation}
\sum_{\lambda_q,\lambda_{\bar q}}
\left|
\mathcal{R}^{00}_{\lambda_q\lambda_{\bar q}}
(x,{\bm k}_\perp)
\right|^2
=
1.
\end{equation}
Evaluating Eq.~\eqref{eq8} explicitly in the LF helicity basis gives\footnote{The diagonal numerators are $(+k^L,+k^R)$ for the present
choice $\bm k_{q\perp}=-\bm k_\perp$ and $(-k^L,-k^R)$ for $\bm k_{q\perp}=+\bm k_\perp$.}
\begin{equation}
\label{Rmatrix}
\mathcal{R}^{00}_{\lambda_q\lambda_{\bar q}}
=
\frac{1}{
\sqrt{2}\sqrt{{\bm k}_\perp^2+\mathcal{A}^2}
}
\begin{pmatrix}
k^L & \mathcal{A}\\
-\mathcal{A} & k^R
\end{pmatrix},
\end{equation}
where
$k^{R(L)}=k_x\pm i k_y$ and
$\mathcal{A}=(1-x)m_q+xm_{\bar q}$.

For the lowest radial state, we employ the Gaussian ansatz
\begin{equation}
\label{eq:9}
\phi(x,{\bm k}_\perp)
=
\frac{4\pi^{3/4}}{\beta^{3/2}}
\sqrt{\frac{\partial k_z}{\partial x}}\,
\exp\left(-\frac{{\vec k}^{\,2}}{2\beta^2}\right),
\end{equation}
where ${\vec k}=({\bm k}_\perp,k_z)$.
The parameter $\beta$ controls the momentum-space width of the wave
function and is fixed phenomenologically from the meson spectrum.
The longitudinal momentum $k_z$ is related to the LF variables by
\begin{equation}
k_z
=
\left(x-\frac{1}{2}\right)M_0
+
\frac{m_{\bar q}^2-m_q^2}{2M_0}.
\end{equation}
The corresponding Jacobian is
\begin{equation}
\frac{\partial k_z}{\partial x}
=
\frac{M_0}{4x(1-x)}
\left[
1-
\left(
\frac{m_q^2-m_{\bar q}^2}{M_0^2}
\right)^2
\right].
\label{eq:radial-jacobian-label}
\end{equation}
This Jacobian is included explicitly in Eq.~\eqref{eq:9} so that the
Gaussian wave function retains the conventional three-dimensional
momentum-space normalization.

The radial wave function is normalized according to
\begin{equation}
\label{eq:10}
\int_0^1\mathrm dx
\int\frac{\mathrm d^2{\bm k}_\perp}{2(2\pi)^3}
\left|
\phi(x,{\bm k}_\perp)
\right|^2
=
1.
\end{equation}
Together with the unit normalization of the spin--orbit wave function, this
radial normalization ensures that the complete LFWF satisfies
Eq.~\eqref{eq:Psinorm} and is therefore consistent with the meson-state
normalization in Eq.~\eqref{eq:statenorm}.

\subsection{Helicity kernels and matrix elements}

We consider the pseudoscalar-to-pseudoscalar transition
$P(q_1\bar q)\to P'(q_2\bar q)$ in the DYW frame. The active quark
changes flavor from $q_1$ to $q_2$, while the antiquark $\bar q$
acts as a spectator. Consistently with the internal momentum convention
introduced above, the constituent momenta in the initial and final
mesons are assigned as
\begin{alignat}{2}
p_1^+ &= xP^+,
&\qquad
p_2^+ &= xP'^+=xP^+,
\nonumber\\
p_{\bar q}^+ &= (1-x)P^+,
&\qquad
p_{\bar q}^{\prime +}
&= (1-x)P'^+=(1-x)P^+,
\nonumber\\
{\bm p}_{1\perp}
&=x{\bm P}_\perp-{\bm k}_\perp,
&\qquad
{\bm p}_{2\perp}
&=x{\bm P}'_\perp-{\bm k}'_\perp,
\nonumber\\
{\bm p}_{\bar q\perp}
&=(1-x){\bm P}_\perp+{\bm k}_\perp,
&\qquad
{\bm p}_{\bar q\perp}'
&=(1-x){\bm P}'_\perp+{\bm k}'_\perp .
\label{eq:LFmomenta}
\end{alignat}
The spectator condition requires the antiquark momentum to remain
unchanged at the weak vertex,
$p_{\bar q}^{\prime +}=p_{\bar q}^+$ and
${\bm p}_{\bar q\perp}'={\bm p}_{\bar q\perp}$.
Taking $q=P-P'$ and choosing
${\bm P}_\perp={\bm 0}_\perp$, this condition gives
${\bm k}'_\perp={\bm k}_\perp+(1-x){\bm q}_\perp$.

In the BT-based LFQM, the hadronic weak-current matrix element is
represented in the impulse approximation as an overlap of the
initial- and final-state radial wave functions,
\begin{equation}
\langle {\cal J}_W^\mu\rangle_{\rm BT}
=
\int[\mathrm d^3{\bm p}_1]\,
\phi_2(x,{\bm k}'_\perp)
\phi_1(x,{\bm k}_\perp)\,
\mathcal{H}^\mu,
\label{eq:WeakCurrentConvolution}
\end{equation}
where the integration measure is
$[\mathrm d^3{\bm p}_1] =\frac{P^+\,\mathrm dx\,\mathrm d^2{\bm k}_\perp}{2(2\pi)^3}$.
All spin-dependent information is contained in the helicity kernel
$\mathcal{H}^\mu$, defined by
\begin{align}
\mathcal{H}^\mu
={}&
\sum_{\lambda_1,\lambda_2,\bar\lambda}
\mathcal{R}^{00\dagger}_{\lambda_2\bar\lambda}
(x,{\bm k}'_\perp)
\left[
\frac{\bar u_{\lambda_2}(p_2)}{\sqrt{p_2^+}}
\gamma^\mu
\frac{u_{\lambda_1}(p_1)}{\sqrt{p_1^+}}
\right]
\mathcal{R}^{00}_{\lambda_1\bar\lambda}
(x,{\bm k}_\perp)
\nonumber\\
\equiv{}&
\sum_{\lambda_1,\lambda_2,\bar\lambda}
\mathcal{H}^\mu_{\lambda_1\bar\lambda
\to\lambda_2\bar\lambda}.
\end{align}
Here, $\lambda_1$ and $\lambda_2$ label the LF helicities of the
initial and final active quarks, respectively, and
$\bar\lambda\equiv\lambda_{\bar q}$ denotes the helicity of the
spectator antiquark.

We separate the active-quark helicity transitions into nonflip (NF)
and helicity-flip (HF) sectors,
\begin{equation}
\mathcal{H}^\mu
=
\mathcal{H}^\mu_{\rm NF}
+
\mathcal{H}^\mu_{\rm HF}.
\end{equation}
The two contributions are
\begin{align}
\mathcal{H}^\mu_{\rm NF}
={}&
\sum_{\bar\lambda}
\left(
\mathcal{H}^\mu_{\uparrow\bar\lambda
\to\uparrow\bar\lambda}
+
\mathcal{H}^\mu_{\downarrow\bar\lambda
\to\downarrow\bar\lambda}
\right),
\\
\mathcal{H}^\mu_{\rm HF}
={}&
\sum_{\bar\lambda}
\left(
\mathcal{H}^\mu_{\uparrow\bar\lambda
\to\downarrow\bar\lambda}
+
\mathcal{H}^\mu_{\downarrow\bar\lambda
\to\uparrow\bar\lambda}
\right).
\end{align}
Substituting the pseudoscalar spin--orbit wave functions and the LF
helicity-spinor matrix elements of Table~III in Ref.~\cite{LB80}, we
obtain the expressions below. For the transverse current we display
the contractions with ${\bm q}_\perp$, which are the quantities entering
$f_-^{(\perp)}$. Components of ${\bm{\mathcal H}}^\perp$ orthogonal
to ${\bm q}_\perp$ do not enter $f_-^{(\perp)}$ and are therefore
not displayed. The helicity-nonflip contributions are
\begin{align}
\mathcal H^+_{\rm NF}&=\frac{2\mathcal W}{\mathcal D},
\nonumber\\
{\bm{\mathcal H}}^\perp_{\rm NF}\cdot{\bm q}_\perp
&=-\frac{\mathcal W(2{\bm k}_\perp\cdot{\bm q}_\perp+{\bm q}_\perp^2)}
{xP^+\mathcal D},
\nonumber\\
\mathcal H^-_{\rm NF}&=\frac{2}{x^2(P^+)^2\mathcal D}
\left[\mathcal W({\bm k}_\perp^2+{\bm k}_\perp\cdot{\bm q}_\perp+m_1m_2)
+(1-x)({\bm k}_\perp\times{\bm q}_\perp)^2\right].
\label{eq:helicity-NF}
\end{align}
The helicity-flip contributions are
\begin{align}
\mathcal H^+_{\rm HF}&=0,
\nonumber\\
{\bm{\mathcal H}}^\perp_{\rm HF}\cdot{\bm q}_\perp
&=\frac{(m_1-m_2)(\mathcal A_2{\bm k}_\perp\cdot{\bm q}_\perp
-\mathcal A_1{\bm k}'_\perp\cdot{\bm q}_\perp)}{xP^+\mathcal D},
\nonumber\\
\mathcal H^-_{\rm HF}
&=\frac{1}{x^2(P^+)^2\mathcal D}\biggl[
2(m_2-m_1)(\mathcal A_2-\mathcal A_1){\bm k}_\perp^2
+2m_1\mathcal A_1(1-x){\bm q}_\perp^2
\nonumber\\[-0.3ex]
&\hspace{3cm}-2({\bm k}_\perp\cdot{\bm q}_\perp)
\{m_1(\mathcal A_2-\mathcal A_1)+(m_2-m_1)\mathcal A_1(1-x)\}\biggr].
\label{eq:helicity-HF}
\end{align}
Here, for compactness, we have defined
$\mathcal D=\sqrt{\mathcal A_1^2+{\bm k}_\perp^2}
\sqrt{\mathcal A_2^2+{\bm k}_\perp^{\prime\,2}}$ and
$\mathcal W={\bm k}_\perp\cdot{\bm k}'_\perp
+\mathcal A_1\mathcal A_2$.
The plus component is entirely helicity nonflip, while the transverse
contraction and minus component receive contributions from both
helicity sectors. These results coincide with the corresponding
contractions obtained from the on-shell trace
$S^\mu_{\rm on}$ in Eq.~\eqref{eq:315}.

\subsection{On-shell overlaps and the partially BT-based prescription}

The determination of $f_-(q^2)$ requires the plus component together
with either the transverse or the minus component. Although only the
$(+,\perp)$ extraction was considered in the covariant BS model, we
also examine the $(+,-)$ current-component extraction in the BT-based
LFQM. We denote the result obtained from
$(\mathcal{J}_W^+,{\bm{\mathcal J}}_W^\perp)$ by
$f_-^{(\perp)}(q^2)$ and that obtained from
$(\mathcal{J}_W^+,\mathcal{J}_W^-)$ by $f_-^{(-)}(q^2)$.

In the BT-based LFQM, the on-shell two-body overlaps give the
plus- and transverse-current form-factor extractions in the
$q^+=0$ frame as
\begin{equation}
\label{eq:fBTplusperpval}
f_+(q^2)
=
\frac{\langle \mathcal{J}_W^+ \rangle_{\rm BT}}
{\mathcal{P}^+},
\qquad
f_-^{(\perp)}(q^2)
=
\frac{\langle \mathcal{J}_W^+ \rangle_{\rm BT}}
{\mathcal{P}^+}
+
\frac{
\langle {\bm{\mathcal J}}_W^\perp \rangle_{\rm BT}
\cdot {\bm q}_\perp
}{
{\bm q}_\perp^2
}.
\end{equation}
For the extraction based on the plus and minus current components,
we define the physical-mass pBT baseline by
\be\label{eq:fBTminuscurrentval}
[f_-^{(-)}(q^2)]_{\rm pBT}
=
\frac{\langle \mathcal{J}_W^- \rangle_{\rm BT}}{q^-}
-
\frac{\mathcal{P}^-}{q^-}
\frac{\langle \mathcal{J}_W^+ \rangle_{\rm BT}}
{\mathcal{P}^+}.
\ee

Each form factor can be written as an overlap of the radial wave
functions with the corresponding projected current kernel,
\begin{align}
f_+(q^2)
&=\int[\mathrm d^3{\bm p}_1]\,
\phi_2(x,{\bm k}'_\perp)\phi_1(x,{\bm k}_\perp)\,\mathscr K_+,
\label{fppval}\\
f_-^{(\perp)}(q^2)
&=\int[\mathrm d^3{\bm p}_1]\,
\phi_2(x,{\bm k}'_\perp)\phi_1(x,{\bm k}_\perp)\,\mathscr K_\perp,
\label{eq:fperpval}\\
[f_-^{(-)}(q^2)]_{\rm pBT}
&=\int[\mathrm d^3{\bm p}_1]\,
\phi_2(x,{\bm k}'_\perp)\phi_1(x,{\bm k}_\perp)\,\mathscr K_-^M,
\label{eq:fminusval}
\end{align}
with
\begin{equation}
\mathscr K_+=\frac{\mathcal H^+}{\mathcal P^+},\qquad
\mathscr K_\perp=\mathscr K_+
+\frac{{\bm{\mathcal H}}^\perp\cdot{\bm q}_\perp}{{\bm q}_\perp^2},\qquad
\mathscr K_-^M=\frac{\mathcal H^--\mathcal P^-\mathscr K_+}{q^-}.
\label{eq:BT-projector-kernels}
\end{equation}
The plus and transverse projections contain no mass-dependent
Lorentz factors involving $M$ or $M'$. The transverse extraction
$f_-^{(\perp)}$ is therefore identical in pBT and fBT and provides
a common reference for testing the minus-current projection,
which contains both $q^-$ and $\mathcal P^-$. Within the present
BT realization, the $(+,\perp)$ extraction is zero-mode free and
is saturated by the on-shell two-body overlap. We therefore do
not attach a ``val'' label to $f_+(q^2)$ or
$f_-^{(\perp)}(q^2)$.

The pBT baseline in Eq.~\eqref{eq:fminusval} combines the
on-shell current overlap with the physical values of both $q^-$ and
$\mathcal P^-$.

This baseline provides a clean internal test of the BT organization.
The full BT prescription modifies only the internal minus-current
projection while leaving the radial wave functions and the elementary
one-body current unchanged. The difference between the two
prescriptions can therefore be traced to the treatment of the
interaction-dependent mass mismatch rather than to different model
parameters or a modified current operator.

\subsection{Full BT prescription and current-component independence}

The fBT prescription extends the constituent-level organization of the
current to the internal Lorentz projection. We replace
$M\to M_0(x,{\bm k}_\perp)$ and
$M'\to M_0'(x,{\bm k}'_\perp)$ in the mass-dependent Lorentz projection factors
under the overlap integral. This is a specification of the current
extraction in the adopted one-body approximation, not a replacement of
the physical external momenta. Unlike an a posteriori correction to an
integrated form factor, it acts on every constituent configuration and
introduces no additional adjustable strength.

For the extraction based on the plus and minus current components, the
fBT projection may be written schematically as
\begin{equation}
\label{eq:fBTminuscurrent}
[f_-^{(-)}(q^2)]_{\rm fBT}
=
\left\langle
\frac{{\cal J}_W^-}{q_{\rm on}^-}
-
\frac{{\cal P}_{\rm on}^-}{q_{\rm on}^-}
\frac{{\cal J}_W^+}{{\cal P}^+}
\right\rangle_{\rm BT},
\end{equation}
with $q_{\rm on}^-$ and ${\cal P}_{\rm on}^-$ given in
Eq.~\eqref{eq:BT-free-minus-factors}.
More explicitly, the fBT result is defined by the LFWF convolution
\begin{equation}
\label{eq:fminusfull}
[f_-^{(-)}(q^2)]_{\rm fBT}
=
\int[\mathrm d^3{\bm p}_1]\,
\phi_2(x,{\bm k}'_\perp)
\phi_1(x,{\bm k}_\perp)\,
\mathscr K_-^{\rm on},
\end{equation}
where
\be\label{eq:Kon-}
\mathscr K_-^{\rm on}=\frac{\mathcal H^--{\cal P}_{\rm on}^-\mathscr K_+}{q_{\rm on}^-}.
\ee
The factors $q_{\rm on}^-$ and ${\cal P}_{\rm on}^-$ depend on the integration
variables. Thus Eq.~\eqref{eq:fBTminuscurrent} is shorthand for the
projected convolution in Eq.~\eqref{eq:fminusfull}, not division of an
already integrated matrix element by a mean invariant mass.
For spacelike transfer, the finite projection follows from the
internal-orientation cancellation summarized in
Appendix~\ref{app:radial-identity}.

We define the BT completion relative to the pBT baseline by
\begin{equation}
\Delta f_{-,\rm BT}^{(-)}(q^2)
\equiv[f_-^{(-)}(q^2)]_{\rm fBT}
     -[f_-^{(-)}(q^2)]_{\rm pBT}.
\label{eq:fBTzmdefinition}
\end{equation}
The completion is already incorporated in the self-consistent
LFQM result of Ref.~\cite{BDchoi}, which implements the
constituent-mass projection.

At the kernel level, subtracting $\mathscr K_-^M$ in
Eq.~\eqref{eq:BT-projector-kernels} from $\mathscr K_-^{\rm on}$
in Eq.~\eqref{eq:Kon-} gives the exact projector identity
\begin{align}
\mathscr K_-^{\rm on}-\mathscr K_-^M
&=\frac{q^- -q_{\rm on}^-}{q^-}\,\mathscr K_-^{\rm on}
  +\frac{\mathcal P^- -{\cal P}_{\rm on}^-}{q^-}\,\mathscr K_+
\nonumber\\
&=\frac{(\delta M_1^2-\delta M_2^2)\mathscr K_-^{\rm on}
+(\delta M_1^2+\delta M_2^2)\mathscr K_+}
{M^2-M'^2-{\bm q}_\perp^2},
\label{eq:BT-projector-difference}
\end{align}
where $\delta M_1^2=M^2-M_0^2$ and $\delta M_2^2=M'^2-M_0'^2$.
The factor $q^- -q_{\rm on}^-$ is the LF energy mismatch in
the constituent-current identity of Eq.~\eqref{eq:constituent-energy-identity}.
The overlap of Eq.~\eqref{eq:BT-projector-difference} is precisely
$\Delta f_{-,\rm BT}^{(-)}$. This identity shows that the BT
completion corrects the mismatch between the physical and
constituent mass combinations in the internal projection.
Its value is fixed once the constituent masses, wave functions,
and pBT baseline are specified. Its separation from the complete
form factor remains baseline dependent, so it is neither a
separate observable nor a model-independent interaction-current
correction.

The dynamical meaning of the mass differences follows from the BT
operator equation. Since $\hat M=\hat M_0+\hat V_{q\bar q}$,
\begin{equation}
 \hat M^2-\hat M_0^2
 =\hat M_0\hat V_{q\bar q}+\hat V_{q\bar q}\hat M_0
  +\hat V_{q\bar q}^{\,2}.
 \label{eq:BT-interaction-square}
\end{equation}
In the internal momentum representation, the eigenvalue equation
therefore relates $(M^2-M_0^2)\Psi$ to the action of the
interaction-dependent terms on the right-hand side. The mismatch
thus reflects binding dynamics rather than a pointwise identification
with $\hat V_{q\bar q}$. For free mass eigenstates the differences vanish
on their support and the completion is absent.

The nontrivial integrated consistency relation is
\begin{equation}
\label{eq:ccind}
f_-^{\rm fBT}(q^2)
=f_-^{(\perp)}(q^2)
=[f_-^{(-)}(q^2)]_{\rm fBT}.
\end{equation}
Unlike Eq.~\eqref{eq:BT-projector-difference}, Eq.~\eqref{eq:ccind}
is a statement about the integrated overlaps. We measure it by
the integrated residual
\begin{equation}
\mathscr R(q^2)\equiv
[f_-^{(-)}(q^2)]_{\rm fBT}-f_-^{(\perp)}(q^2)
=\int[\mathrm d^3{\bm p}_1]\,\phi_2\phi_1
\left(\mathscr K_-^{\rm on}-\mathscr K_\perp\right).
\label{eq:BT-integrated-residual}
\end{equation}

The kernels differ point by point. For
$\phi_i=\sqrt{\partial k_{zi}/\partial x}\,u_i({\bm k}_{i\perp}^{\,2}+k_{zi}^2)$
with the specified pseudoscalar spin coupling, the radial factors
separate from the internal orientation at fixed $(M_0^2,M_0'^2)$.
Appendix~\ref{app:radial-identity} shows that the remaining difference
is the vanishing boundary term in
Eq.~\eqref{eq:orientation-boundary-zero}. Thus $\mathscr R(q^2)=0$
is independent of the radial shape within this convergent class.
It extends to the timelike region wherever both continued overlaps
are regular on the same analytic branch.

Combining Eqs.~\eqref{eq:fBTzmdefinition} and \eqref{eq:ccind}
gives $\Delta f_{-,\rm BT}^{(-)}
=f_-^{(\perp)}-[f_-^{(-)}]_{\rm pBT}$.
This is the exact integrated correction required for the pBT
minus-current extraction to reproduce the common transverse result.
Current-component independence thus identifies the effective
zero-mode contribution missing from the pBT extraction within
this valence-only realization.

The pBT baseline also has a kinematic limitation. Its minus-current
projection contains the physical factor
$q^-=(M^2-M'^2+q^2)/P^+$ and therefore becomes singular at
$q^2=-(M^2-M'^2)$ unless the corresponding numerator vanishes.
The on-shell BT overlap does not in general satisfy this
physical-mass cancellation. The pBT contribution and the BT
completion can therefore develop opposite kinematic poles that
cancel in their sum, while the fBT form factor remains regular.
This behavior further shows that the pBT/completion split is a
baseline-dependent diagnostic rather than a separation into physical
contributions.

To connect the internal projection back to physical hadron kinematics,
write the completed matrix element as
\begin{equation}
\langle\mathcal J_W^\mu\rangle_{\rm comp}
=\mathcal P^\mu f_+(q^2)+q^\mu f_-^{\rm fBT}(q^2).
\label{eq:BT-completed-current}
\end{equation}
The external four-vectors and masses in this expression are physical.
Using the pBT definition, its minus-component difference from the
uncompleted on-shell overlap is
\begin{equation}
\langle\mathcal J_W^-\rangle_{\rm comp}
-\langle\mathcal J_W^-\rangle_{\rm BT}
=q^-\Delta f_{-,\rm BT}^{(-)}.
\label{eq:BT-completed-minus}
\end{equation}
Equations~\eqref{eq:BT-projector-difference} and
\eqref{eq:BT-completed-minus} make the role of fBT explicit. The
constituent-level projection supplies a definite effective completion
while the reconstructed hadron matrix element retains its physical
Lorentz decomposition.
This reconstruction does not establish a unique microscopic
two-body current.

\subsection{Jaus's covariant extension and its practical LFQM implementation}
\label{sec:jaus}

As discussed in Sec.~III, Jaus's covariant extension incorporates
missing zero-mode contributions through effective replacements that
remove the corresponding $C$ coefficients. Strict covariance also
requires the integrated identities involving the residual
$B$ coefficients~\cite{Jaus99}. For pseudoscalar-to-pseudoscalar
transitions, this construction gives the zero-mode-free form factor
$F_1=f_+$ and the zero-mode-completed form factor $F_2=f_-$.

For phenomenological applications, Jaus evaluated the covariantly
derived form-factor expressions using the standard symmetric LF
vertex functions of Eqs.~(5.2)--(5.4) in Ref.~\cite{Jaus99}. These
vertices provide substantially better light-meson phenomenology than
the asymmetric vertex functions required by the strict covariant
construction. Jaus noted, however, that the standard symmetric
vertices do not satisfy the covariance identities [see Eq.~(3.32) of Ref.~\cite{Jaus99}].
Their practical use therefore leaves residual dependence on the LF
orientation through the $B$ coefficients.

To compare this standard-vertex implementation with the fBT
projection using the same radial wave functions, we specify the
mapping at the level of the product of the two vertices,
\begin{equation}
\label{eq:BS-LF-map}
\frac{
\Nraw\,\chi_1(x,{\bm k}_\perp)
\chi_2(x,{\bm k}'_\perp)
}{1-x}
\ \longmapsto\
\frac{1-x}{2}\,
\frac{
\phi_1(x,{\bm k}_\perp)
\phi_2(x,{\bm k}'_\perp)
}{
\sqrt{\mathcal A_1^2+{\bm k}_\perp^2}
\sqrt{\mathcal A_2^2+{\bm k}_\perp^{\prime\,2}}
}.
\end{equation}
The arrow denotes a model correspondence rather than an equality
between vertex functions belonging to different dynamical
descriptions. Specifying the correspondence for the vertex product
also avoids assigning a common transition normalization separately
to the two vertices. In what follows, $f_-^{\rm Jaus}$ denotes only
this practical standard-vertex implementation.

Applying Eq.~\eqref{eq:BS-LF-map} to the zero-mode-completed
covariant expression in Eqs.~\eqref{eq:fmconv} and
\eqref{eq:fmspinfull} gives the standard-vertex result corresponding
to Eq.~(4.3) of Ref.~\cite{Jaus99},
\begin{equation}
\label{eq:Jausf-}
f_-^{\rm Jaus}(q^2)
=
\frac{1}{2}
\int_0^1(1-x)\,\mathrm dx
\int\frac{\mathrm d^2{\bm k}_\perp}{16\pi^3}\,
\frac{\phi_1(x,{\bm k}_\perp)}
{\sqrt{{\cal A}_1^2+{\bm k}_\perp^2}}
\frac{\phi_2(x,{\bm k}'_\perp)}
{\sqrt{{\cal A}_2^2+{\bm k}_\perp^{\prime\,2}}}
\left[
\frac{S^+_{\rm val}}{\mathcal P^+}
+
\frac{
\left(
S^\perp_{\rm val}
+
S^\perp_{{\rm Z.M.},{\rm eff}}
\right)
\cdot{\bm q}_\perp
}{
{\bm q}_\perp^2
}
\right].
\end{equation}
Equation~\eqref{eq:Jausf-} transfers the BS valence current
together with the effective zero-mode operator of
Eq.~\eqref{eq:SZMop} to the Gaussian LFQM through the mapping in
Eq.~\eqref{eq:BS-LF-map}.
This is the practical standard-vertex implementation considered
here. Jaus formulated this construction in terms of covariant and
standard LF vertex functions rather than in BT terminology.

The comparison with fBT is particularly instructive for the
$(+,\perp)$ current-component extraction. In the covariant BS model,
$f_-^{(\perp)}(q^2)$ requires an LF zero-mode contribution. In the
BT-based LFQM considered here, the same extraction is zero-mode free
and the on-shell two-body overlap gives the complete result. Using
the same radial wave functions as in Eq.~\eqref{eq:Jausf-}, the fBT
result is
\begin{equation}
\label{eq:fBTf-}
f_-^{(\perp)}(q^2)
=
\frac{1}{2}
\int_0^1(1-x)\,\mathrm dx
\int\frac{\mathrm d^2{\bm k}_\perp}{16\pi^3}\,
\frac{\phi_1(x,{\bm k}_\perp)}
{\sqrt{{\cal A}_1^2+{\bm k}_\perp^2}}
\frac{\phi_2(x,{\bm k}'_\perp)}
{\sqrt{{\cal A}_2^2+{\bm k}_\perp^{\prime\,2}}}
\left[
\frac{S^+_{\rm on}}{\mathcal P^+}
+
\frac{
S^\perp_{\rm on}\cdot{\bm q}_\perp
}{
{\bm q}_\perp^2
}
\right].
\end{equation}

The difference between the two prescriptions follows directly by
subtracting Eqs.~\eqref{eq:Jausf-} and \eqref{eq:fBTf-}.
Since $S^+_{\rm val}=S^+_{\rm on}$, the plus-current terms cancel.
At the spectator pole,
$S^\perp_{\rm val}=S^\perp_{\rm on}+S^\perp_{\rm inst}$.
We therefore obtain
\begin{align}
f_-^{\rm Jaus}-f_-^{\rm fBT}
={}&
\frac12
\int_0^1(1-x)\,\mathrm dx
\int\frac{\mathrm d^2{\bm k}_\perp}{16\pi^3}
\frac{\phi_1\phi_2}{\mathcal D}
\frac{
\left(
S^\perp_{\rm inst}
+
S^\perp_{{\rm Z.M.},{\rm eff}}
\right)
\cdot{\bm q}_\perp
}{
{\bm q}_\perp^2
}.
\label{eq:mapped-instantaneous-and-ZM}
\end{align}
Defining the mapped instantaneous and zero-mode contributions as
$I_{\rm inst}$ and $I_{\rm ZM}$, respectively, we have
$f_-^{\rm Jaus}-f_-^{\rm fBT}=I_{\rm inst}+I_{\rm ZM}$.
We define the difference between the fBT and practical Jaus results by
\begin{equation}
\label{eq:Jaus-BT-difference}
\Delta f_{-,{\rm BT-Jaus}}
\equiv
f_-^{\rm fBT}-f_-^{\rm Jaus}
=
-\left(I_{\rm inst}+I_{\rm Z.M.}\right).
\end{equation}
The difference in Eq.~\eqref{eq:Jaus-BT-difference} is
distinct from the BT completion defined in
Eq.~\eqref{eq:fBTzmdefinition}. In the present BT-based LFQM,
the on-shell transverse overlap already reproduces the independent
fBT minus-current result for $f_-$ in Eq.~\eqref{eq:ccind}.
The standard-vertex implementation of Jaus's prescription
retains the mapped instantaneous and zero-mode terms in addition
to this on-shell contribution.

\subsection{Direct analytic continuation of the BT overlaps}
\label{sec:direct-continuation}

The timelike form factors are obtained by evaluating the analytically
continued $q^+=0$ overlaps directly on a complex transverse contour,
rather than by fitting or extrapolating their spacelike values.
Writing $t=q^2>0$, we use ${\bm\ell}_\perp\in\mathbb R^2$ as
the integration variable and choose the mass fraction
$\eta=m_1/(m_1+m_2)$ to distribute the imaginary transverse
displacement between the initial and final wave functions according to
\begin{equation}
\begin{aligned}
{\bm q}_\perp
&=(i\sqrt{t},0),\\
{\bm k}_\perp
&={\bm\ell}_\perp-\eta(1-x){\bm q}_\perp,\\
{\bm k}'_\perp
&={\bm\ell}_\perp+(1-\eta)(1-x){\bm q}_\perp.
\end{aligned}
\label{eq:continued-contour}
\end{equation}
This mass-weighted contour preserves the spectator relation
${\bm k}'_\perp-{\bm k}_\perp=(1-x){\bm q}_\perp$.
The original Gaussian wave functions and current kernels are evaluated
with these analytically continued momenta without introducing a new
parametrization. Scalar products retain their bilinear analytic form.
The square-root branches are fixed by continuity from the
spacelike region, and the continued radial wave functions enter
without complex conjugation.

For the Gaussian wave functions considered here, the conditions
$\eta^2t<m_1^2$ and $(1-\eta)^2t<m_2^2$ suffice to avoid the
relevant wave-function singularities.  For a general displacement
fraction $0<\eta<1$, the common sufficient domain is
$t<\min[m_1^2/\eta^2,m_2^2/(1-\eta)^2]$. The first bound decreases
with increasing $\eta$, while the second increases.
The chosen mass fraction makes the two bounds equal and thus
gives the largest interval guaranteed by both conditions,
$0<t<(m_1+m_2)^2$.

For a $P\to P'$ transition, this wave-function analyticity
domain contains the interval $0<t\leq t_{\rm max}$ if
\begin{equation}
t_{\rm max}=(M-M')^2<(m_1+m_2)^2.
\label{eq:physical-continuation-condition}
\end{equation}
The $D\to K$ parameters satisfy this wave-function condition.
Appendix~\ref{app:direct-continuation} checks the square-root
branches and current-projector denominators and establishes
absolute convergence of the full Gaussian overlaps on this
contour for $0<t\leq t_{\rm max}$.

\subsection{$D\to K\ell\nu$ as an illustration of the BT construction}
\label{sec:numerical}

For the semileptonic $D\to K\ell\nu$ decay in the BT-based LFQM, we
use the constituent-quark masses $m_c=1.8$~GeV,
$m_s=0.45$~GeV, and $m_{u(d)}=0.22$~GeV, together with the Gaussian
parameters $\beta_{c\bar q}=0.4679$~GeV and
$\beta_{s\bar q}=0.3886$~GeV~\cite{CJ99b}.
These inputs differ from those of the BS benchmark in Sec.~\ref{sec:bs-benchmark}, 
$(m_c,m_s,m_{u(d)})=(1.60,0.50,0.30)$~GeV and $(\Lambda_D,\Lambda_K)=(1.80,0.90)$~GeV. 
The BS benchmark and the BT calculation thus use different input
parameters, so their numerical difference does not isolate a
zero-mode effect.
The physical masses are $M_D=1.869$~GeV and $M_K=0.494$~GeV, which
give $q^2_{\rm max}=1.8906~\mathrm{GeV}^2$.

\begin{figure}
\centering
\includegraphics[width=3in,height=2.3in]{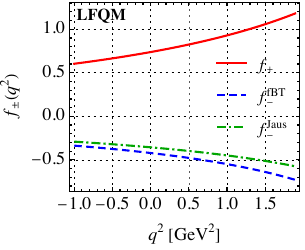}\hspace{0.5cm}
\includegraphics[width=3in,height=2.3in]{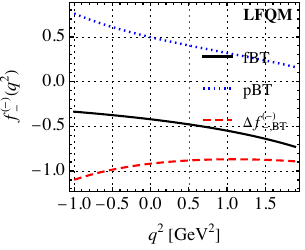}
\caption{Illustration of the BT organization for $D\to K$.
Left: $f_+$, the component-independent $f_-^{\rm fBT}$, and the
practical standard-vertex result $f_-^{\rm Jaus}$ of
Eq.~\eqref{eq:Jausf-}. Right: the minus-current fBT result, the
physical-mass pBT projection of Eq.~\eqref{eq:fminusval}, and the
BT completion $\Delta f_{-,\rm BT}^{(-)}$ defined in
Eq.~\eqref{eq:fBTzmdefinition}. All curves are obtained from direct
overlaps.}
\label{fig2}
\end{figure}

Using the direct continuation described in
Sec.~\ref{sec:direct-continuation}, the entries in
Table~\ref{tab:endpoints} are evaluated directly at the kinematic
endpoints. The $(+,\perp)$ and $(+,-)$ fBT overlaps are integrated
independently. On the 33-point grid spanning
$-1~\mathrm{GeV}^2\leq q^2\leq q^2_{\rm max}$, including both
physical endpoints $q^2=0$ and $q^2=q^2_{\rm max}$, the largest
absolute difference between the independently integrated
transverse- and minus-current overlaps is $2.5\times10^{-12}$ at
quadrature order 120. For $q^2<0$, the fBT minus-current curve
in Fig.~\ref{fig2} is obtained by direct integration of the
minus-current overlap in the original $(x,{\bm k}_\perp)$ variables,
with the $x$ integration performed first at fixed unshifted
transverse momentum. The invariant-mass variables in
Appendix~\ref{app:radial-identity} are used for the analytic argument
and are not the numerical variables used to generate the curves.
The small residual therefore tests the current-component identity
rather than the uncertainty of the constituent model.

Figure~\ref{fig2} covers
$-1~\mathrm{GeV}^2\leq q^2\leq q^2_{\rm max}$.
The left panel shows that $f_-^{\rm Jaus}$, obtained by
applying the BS prescription unchanged to the Gaussian BT-based
LFQM, fails to reproduce $f_-^{\rm fBT}$ even though the mapped Jaus and fBT calculations use identical
constituent masses and Gaussian wave functions.
As shown in Eq.~\eqref{eq:ccind}, $f_-^{\rm fBT}$ is
the common result of the independent $(+,\perp)$ and fBT $(+,-)$
extractions. The right panel isolates the role of the internal
mass projection. For the parameters used here, the BT completion
is negative and exceeds the positive pBT baseline in magnitude,
so their sum gives a negative $f_-^{\rm fBT}$.
No additional parameter is introduced in this completion.
The magnitude of the BT completion decreases from the spacelike
side, reaches a shallow minimum in the timelike interval, and
then increases toward zero recoil.

Table~\ref{tab:endpoints} gives the form factors and the BT
completion at maximum recoil ($q^2=0$) and zero recoil
($q^2=q^2_{\rm max}$). It also lists separately the mapped
instantaneous and zero-mode contributions that enter the
comparison in the left panel of Fig.~\ref{fig2}.

\begin{table}[H]
\caption{Direct-overlap endpoint values for $D\to K$ using the same
LFQM parameters. The zero-recoil point is
$q^2_{\rm max}=1.8906~\mathrm{GeV}^2$. The two fBT extractions are
computed independently. The BT completion is defined relative to pBT.
The mapped contributions satisfy
$\Delta f_{-,{\rm BT-Jaus}}=-(I_{\rm inst}+I_{\rm Z.M.})$.
Values are rounded to four decimal places.}
\label{tab:endpoints}
\begin{ruledtabular}
{\renewcommand{\arraystretch}{1.18}
\setlength{\tabcolsep}{12pt}
\begin{tabular}{lrr}
Quantity & $q^2=0$ & $q^2=q^2_{\rm max}$\\
\hline
$f_+$ & $0.7384$ & $1.1814$\\
$f_-^{(\perp)}$ & $-0.4216$ & $-0.7294$\\
$[f_-^{(-)}]_{\rm fBT}=f_-^{\rm fBT}$ & $-0.4216$ & $-0.7294$\\
$[f_-^{(-)}]_{\rm pBT}$ & $0.4967$ & $0.1629$\\
$\Delta f_{-,\rm BT}^{(-)}$ & $-0.9183$ & $-0.8923$\\
$f_-^{\rm Jaus}$ & $-0.3562$ & $-0.5740$\\
$I_{\rm inst}$ & $0.0374$ & $0.1061$\\
$I_{\rm Z.M.}$ & $0.0281$ & $0.0493$\\
$\Delta f_{-,{\rm BT-Jaus}}$ & $-0.0654$ & $-0.1554$\\
\end{tabular}}
\end{ruledtabular}
\end{table}

At $q^2=0$, the difference between the fBT result $-0.4216$
and the pBT baseline $0.4967$ gives the BT completion $-0.9183$.
Although this difference is not separately observable, its
magnitude and sign can be traced to the two terms in the projector
identity, Eq.~\eqref{eq:BT-projector-difference}. Integrating them
separately, we obtain $-0.8893$ from the term proportional to
$(\delta M_1^2+\delta M_2^2)\mathscr K_+$ and $-0.0290$ from
the term proportional to
$(\delta M_1^2-\delta M_2^2)\mathscr K_-^{\rm on}$.
Their sum gives the tabulated completion.

To interpret the dominant term, we define the average of a
function $O$ of the integration variables using the $f_+$ integrand
as the weight,
$\langle O\rangle\equiv f_+(0)^{-1}
\int[\mathrm d^3{\bm p}_1]\,
\phi_2\phi_1\,\mathscr K_+\,O$,
with all quantities evaluated at $q^2=0$. Neglecting the smaller
$\mathscr K_-^{\rm on}$ contribution gives
\begin{equation}
\Delta f_{-,\rm BT}^{(-)}(0)\simeq
\frac{\langle\delta M_1^2+\delta M_2^2\rangle}{M^2-M'^2}\,f_+(0).
\label{eq:completion-estimate}
\end{equation}
For the Gaussian $D\to K$ wave functions,
$\langle M_0^2\rangle\simeq6.09~\mathrm{GeV}^2$ and
$\langle M_0'^2\rangle\simeq1.57~\mathrm{GeV}^2$ exceed
$M_D^2=3.49~\mathrm{GeV}^2$ and $M_K^2=0.24~\mathrm{GeV}^2$,
respectively. The weighted mass mismatch,
$\langle\delta M_1^2+\delta M_2^2\rangle\simeq-3.91~\mathrm{GeV}^2$,
is comparable in magnitude to $M_D^2-M_K^2=3.25~\mathrm{GeV}^2$.
With $f_+(0)=0.7384$, Eq.~\eqref{eq:completion-estimate} therefore
gives approximately $-0.89$, accounting for most of the completion.

The dominant term also explains the sign reversal in the
minus-current extraction. The pBT projection subtracts the
plus-current contribution with the physical factor $\mathcal P^-$.
For the present parameters, the larger constituent factor
$\mathcal P_{\rm on}^-$ in fBT increases this subtraction,
changing the positive pBT result into the negative fBT result.
The size and sign of the completion thus reflect the mismatch
between the free and physical masses for the adopted wave
functions.

The discrepancy between $f_-^{\rm Jaus}$ and $f_-^{\rm fBT}$
has a different origin. In the standard-vertex implementation of
Jaus's prescription, the mapped instantaneous contribution
$I_{\rm inst}$ from the BS valence current and the mapped
zero-mode contribution $I_{\rm Z.M.}$ are both positive at each
kinematic endpoint (Table~\ref{tab:endpoints}). Their sum shifts
$f_-^{\rm Jaus}$ away from the on-shell transverse result, which
already agrees with the independently evaluated fBT minus-current
result. Thus, applying the standard-vertex correspondence of
Eq.~\eqref{eq:BS-LF-map} to the vertex product while retaining
the BS valence current and the effective zero-mode operator for the BS
endpoint contribution fails to reproduce the component-independent
BT result for $f_-$ in $D\to K$.

\section{Summary and discussion}
\label{sec:summary}

In the present BT-based LFQM, the interaction enters through
the mass operator and wave functions, while the one-body current
is evaluated with on-shell constituent spinors. 
The fBT prescription uses the free invariant masses $M_0$ and
$M_0'$ in the internal Lorentz factors of the minus-current
projection. For spacelike momentum transfer, we prove that this
prescription yields the same $f_-(q^2)$ from the $(+,\perp)$
and $(+,-)$ extractions after integration. The proof applies to
rotationally invariant radial wave functions with the LF Jacobian
and the adopted pseudoscalar spin coupling, provided the reduced
overlaps converge.
At fixed $M_0^2$ and $M_0'^2$, the orientation integral in the
difference between the two extractions reduces to a boundary term
that vanishes at both endpoints of the orientation interval
[Eq.~\eqref{eq:orientation-boundary-zero}]. This gives a direct
BT derivation of the component independence found numerically in
Ref.~\cite{BDchoi} by mapping the on-shell BS contributions to
the LFQM.

The transverse extraction is common to pBT and fBT, and the
independently evaluated fBT minus-current extraction reproduces
this result [Eq.~\eqref{eq:ccind}]. The fBT--pBT difference
defined in Eq.~\eqref{eq:fBTzmdefinition} is therefore the exact
integrated correction that must be added to the pBT minus-current
result to recover the common transverse result. Within the
valence-only BT-based LFQM, current-component independence
indirectly identifies this correction as the effective zero-mode
contribution missing from the pBT minus-current projection.
The completion is fixed once the model inputs and the pBT baseline
are specified and is already included in the fBT result.
The projector identity in Eq.~\eqref{eq:BT-projector-difference}
expresses its mass dependence in terms of $M^2-M_0^2$ and
$M'^2-M_0'^2$. Equation~\eqref{eq:BT-interaction-square} relates
these differences to the binding interaction, and the completion
vanishes in the free limit.

In the covariant BS benchmark, a finite transverse-current
contribution survives as the nonvalence region shrinks to the
longitudinal endpoint in the limit $q^+\to0^+$. The effective
zero-mode operator reproduces this directly calculated endpoint
contribution within the valence convolution. Supplementing the BS
valence current with this operator therefore recovers the
independently calculated covariant $f_-$. In the BT-based LFQM, the transverse extraction is already
saturated by the on-shell overlap.

To test whether the BS prescription carries over to the
BT-based LFQM, we retain both the BS valence current and the
effective zero-mode operator and replace the covariant BS vertex
factors by Gaussian LFQM wave functions according to the
standard-vertex correspondence. As the left panel of
Fig.~\ref{fig2} shows, the resulting $f_-^{\rm Jaus}$ fails to
reproduce the component-independent $f_-^{\rm fBT}$ even though
both calculations use identical constituent masses and Gaussian
wave functions. The mapped instantaneous part of the BS valence
current and the mapped zero-mode contribution have the same sign
at each kinematic endpoint and thus do not cancel
[Table~\ref{tab:endpoints}]. This result rules out an unchanged
transfer of Jaus's BS prescription to the Gaussian BT realization
considered here. The later type-II
studies~\cite{QC18,ChangLiWang19} modify the BS-to-LFQM correspondence.

Since the fBT and Jaus calculations use the same $f_+$ and
external meson masses, their difference in $f_-$ implies
$\Delta f_0=q^2\Delta f_-/(M^2-M'^2)$. At $q^2=0$, both
prescriptions therefore give $f_0(0)=f_+(0)$ despite their
different values of $f_-(0)$.

We use a mass-weighted complex transverse contour to
analytically continue the Gaussian overlap integrals from the
spacelike region to the timelike region. On this contour,
$0<t<(m_1+m_2)^2$ is sufficient for wave-function analyticity.
When Eq.~\eqref{eq:physical-continuation-condition} holds, this
domain contains the entire interval $0<t\leq t_{\rm max}$,
including zero recoil. Direct evaluation also requires the
current-projector denominators to remain nonzero on the contour
for $t>0$. For $D\to K$, Appendix~\ref{app:direct-continuation}
verifies this requirement and establishes absolute convergence of
the full overlaps for $0<t\leq t_{\rm max}$. The independently evaluated
transverse and fBT minus-current extractions agree numerically
up to zero recoil without a form-factor parametrization or
extrapolation.

\begin{acknowledgments}
This work was supported by the National Research Foundation of Korea (NRF) grant funded by the Korea government
under Grant No. RS-2026-25475296.
\end{acknowledgments}

\appendix
\renewcommand{\theequation}{\Alph{section}.\arabic{equation}}

\section{Covariant Feynman-parameter representation}
\label{app:covariant-detail}

For completeness, we summarize the essential steps connecting the
covariant triangle amplitude in Eq.~\eqref{eq:31} to the form factors
in Eqs.~\eqref{eq:36} and \eqref{eq:37}. The intermediate trace
algebra follows the standard treatment of the solvable covariant BS
model~\cite{BCJ01,BCJ03} and is not repeated here.

The product of the five propagator denominators is first decomposed as
\begin{align}
\label{eq:34}
\frac{1}{
N_{\Lambda_1}N_1N_{\bar q}N_2N_{\Lambda_2}
}
&=
\frac{1}{
(\Lambda_1^2-m_1^2)(\Lambda_2^2-m_2^2)
}
\frac{1}{N_{\bar q}}
\left(
\frac{1}{N_{\Lambda_1}}-\frac{1}{N_1}
\right)
\left(
\frac{1}{N_{\Lambda_2}}-\frac{1}{N_2}
\right).
\end{align}
The original loop integral is thus expressed as an alternating sum of
four terms involving three propagators. For a generic term, we define
$N_a=p_1^2-a^2+i\epsilon$ and
$N_b=p_2^2-b^2+i\epsilon$, with
$a\in\{m_1,\Lambda_1\}$ and
$b\in\{m_2,\Lambda_2\}$.

The three denominators are combined according to
\begin{equation}
\frac{1}{N_aN_{\bar q}N_b}
=
2\int_0^1dx\int_0^{1-x}dy\,
\frac{1}{
\left[(k-K)^2-C_{ab}+i\epsilon\right]^3
},
\end{equation}
where $K^\mu=xP^\mu+yP'^\mu$ and
\begin{equation}
C_{ab}
=
xa^2+yb^2+(1-x-y)m_{\bar q}^2
-(1-x-y)(xM^2+yM'^2)-xyq^2.
\end{equation}
Here, $x$ and $y$ are Feynman parameters in the triangular domain
$0\leq x\leq1$ and $0\leq y\leq1-x$.

After shifting the loop momentum as $\ell^\mu=k^\mu-K^\mu$,
terms odd in $\ell^\mu$ vanish under symmetric integration, while the
quadratic tensor integral is reduced according to
$\ell^\mu\ell^\nu\to g^{\mu\nu}\ell^2/D$.
The remaining loop integrations require
\begin{align}
{\cal I}_0(C)
&\equiv
i\mu^{2\epsilon_D}
\int\frac{d^D\ell}{(2\pi)^D}
\frac{1}{(\ell^2-C+i\epsilon)^3}
=
\frac{1}{32\pi^2C},
\\
{\cal I}_2(C)
&\equiv
i\mu^{2\epsilon_D}
\int\frac{d^D\ell}{(2\pi)^D}
\frac{\ell^2}{(\ell^2-C+i\epsilon)^3}
\nonumber\\
&=
-\frac{1}{16\pi^2}
\left[
{\rm div.}
-\ln\left(\frac{C}{\mu^2}\right)
\right]
+\frac{1}{32\pi^2},
\end{align}
where $D=4-2\epsilon_D$ and ${\rm div.}$ denotes a
mass-independent divergent constant.

In the alternating Pauli--Villars combination generated by
Eq.~\eqref{eq:34}, all mass-independent divergent and
scheme-dependent terms cancel. It is convenient to introduce
\begin{align}
{\cal C}
&=
\frac{1}{C_{\Lambda_1\Lambda_2}}
-\frac{1}{C_{\Lambda_1m_2}}
-\frac{1}{C_{m_1\Lambda_2}}
+\frac{1}{C_{m_1m_2}},
\\
{\cal L}_C
&=
\ln\left(
\frac{
C_{\Lambda_1\Lambda_2}C_{m_1m_2}
}{
C_{\Lambda_1m_2}C_{m_1\Lambda_2}
}
\right).
\end{align}
The finite numerator functions are
\begin{align}
{\cal F}_+(x,y)
&=
(1-x-y)^2(xM^2+yM'^2)
+xy(2-x-y)q^2
\nonumber\\
&\quad
+(x+y)
\left[
m_1m_2-(m_1+m_2)m_{\bar q}
\right]
+m_{\bar q}(m_1+m_2),
\\
{\cal F}_-(x,y)
&=
(x^2-y^2)(xM^2+yM'^2)
+(yM'^2-xM^2)
-xy(x-y)q^2
\nonumber\\
&\quad
+(x-y)
\left[
m_1m_2-(m_1+m_2)m_{\bar q}
\right]
+m_{\bar q}(m_2-m_1).
\end{align}
Combining these numerator functions with the finite
Pauli--Villars combinations ${\cal C}$ and ${\cal L}_C$ reproduces
the covariant form factors in Eqs.~\eqref{eq:36} and
\eqref{eq:37}.

\section{Feynman-parameter representations of the LF valence contributions}
\label{app:valence-reduction}

We summarize the transverse-momentum reduction of the LF valence
contributions in Eq.~\eqref{eq:322}. The plus-current spin factor is
\begin{equation}
\frac{S^+_{\rm val}}{{\cal P}^+}
=
\frac{2}{1-x}
\left(
{\bm k}_\perp\cdot{\bm k}'_\perp
+
{\cal A}_1{\cal A}_2
\right),
\end{equation}
where
${\bm k}'_\perp={\bm k}_\perp+(1-x){\bm q}_\perp$.
The LF vertex denominators may be written as
\begin{align}
\label{eq:326}
\Delta_a
&=
xm_{\bar q}^2+(1-x)a^2-x(1-x)M^2,
\qquad
a\in\{m_1,\Lambda_1\},
\\
\Delta'_b
&=
xm_{\bar q}^2+(1-x)b^2-x(1-x)M'^2,
\qquad
b\in\{m_2,\Lambda_2\}.
\end{align}
The product of the two LF vertex functions then takes the
Pauli--Villars form
\begin{align}
\label{eq:327}
\chi_1(x,{\bm k}_\perp)
\chi_2(x,{\bm k}'_\perp)
&=
\frac{(1-x)^2}{
(\Lambda_1^2-m_1^2)(\Lambda_2^2-m_2^2)
}
\left[
\frac{1}{{\bm k}_\perp^2+\Delta_{m_1}}
-
\frac{1}{{\bm k}_\perp^2+\Delta_{\Lambda_1}}
\right]
\nonumber\\
&\quad\times
\left[
\frac{1}{{\bm k}_\perp^{\prime\,2}+\Delta'_{m_2}}
-
\frac{1}{{\bm k}_\perp^{\prime\,2}+\Delta'_{\Lambda_2}}
\right].
\end{align}

For each of the four terms in Eq.~\eqref{eq:327}, the two transverse
denominators are combined according to
\begin{equation}
\label{eq:328}
\frac{1}{
({\bm k}_\perp^2+\Delta_a)
({\bm k}_\perp^{\prime\,2}+\Delta'_b)
}
=
\int_0^1dy\,
\frac{1}{
({\bm\ell}_\perp^2+\Delta_{ab})^2
},
\end{equation}
where
${\bm\ell}_\perp
={\bm k}_\perp+y(1-x){\bm q}_\perp$ and
\begin{equation}
\label{eq:329}
\Delta_{ab}
=
(1-y)\Delta_a
+y\Delta'_b
+y(1-y)(1-x)^2{\bm q}_\perp^2.
\end{equation}
Terms odd in ${\bm\ell}_\perp$ vanish under transverse integration.
The required transverse integrals are
\begin{equation}
\label{eq:329c}
\int
\frac{d^2{\bm\ell}_\perp}{
({\bm\ell}_\perp^2+\Delta)^2
}
=
\frac{\pi}{\Delta},
\qquad
\int d^2{\bm\ell}_\perp\,
\frac{{\bm\ell}_\perp^2}{
({\bm\ell}_\perp^2+\Delta)^2
}
=
\pi
\left[
{\rm div.}
-1
-\ln\left(\frac{\Delta}{\mu^2}\right)
\right].
\end{equation}
The divergent and mass-independent terms cancel in the alternating
Pauli--Villars combination. Defining
\begin{equation}
\label{eq:RPV}
{\cal R}_{\rm PV}
=
\frac{1}{\Delta_{m_1m_2}}
-\frac{1}{\Delta_{m_1\Lambda_2}}
-\frac{1}{\Delta_{\Lambda_1m_2}}
+\frac{1}{\Delta_{\Lambda_1\Lambda_2}},
\end{equation}
we obtain
\begin{align}
\label{eq:LFcovfplus}
\left[f_+(q^2)\right]_{\rm val}
&=
\frac{\Npv}{8\pi^2}
\int_0^1dx\int_0^1dy
\Biggl\{
\ln\left(
\frac{
\Delta_{\Lambda_1m_2}\Delta_{m_1\Lambda_2}
}{
\Delta_{\Lambda_1\Lambda_2}\Delta_{m_1m_2}
}
\right)
\nonumber\\
&\qquad\qquad
+
\left[
{\cal A}_1{\cal A}_2
-y(1-y)(1-x)^2{\bm q}_\perp^2
\right]
{\cal R}_{\rm PV}
\Biggr\}.
\end{align}

For the $(+,\perp)$ extraction of $f_-(q^2)$,
Eq.~\eqref{eq:328} gives
${\bm k}_\perp\cdot{\bm q}_\perp/q^2
\to y(1-x)$ after transverse integration, where
$q^2=-{\bm q}_\perp^2$ in the DYW frame.
The resulting numerator is
\begin{align}
\label{eq:fmnum}
\widetilde{\cal F}_-(x,y)
&=
-(1-x)M^2
+(m_2-m_{\bar q}){\cal A}_1
-m_{\bar q}(m_1-m_{\bar q})
\nonumber\\
&\quad
+y(1-x)
\left[
M^2+M'^2
-2(m_1-m_{\bar q})(m_2-m_{\bar q})
\right].
\end{align}
Since this numerator contains no term proportional to
${\bm\ell}_\perp^2$, only the first integral in
Eq.~\eqref{eq:329c} contributes. We therefore obtain
\begin{equation}
\label{eq:LFcovfminus}
\left[f_-^{(\perp)}(q^2)\right]_{\rm val}
=
\frac{\Npv}{8\pi^2}
\int_0^1(1-x)\,dx
\int_0^1dy\,
\widetilde{\cal F}_-(x,y)
{\cal R}_{\rm PV}.
\end{equation}
Equations~\eqref{eq:LFcovfplus} and \eqref{eq:LFcovfminus}
are the Feynman-parameter representations of the direct LF valence
convolutions in Eq.~\eqref{eq:322}.

\section{Direct nonvalence zero-mode integral}
\label{app:direct-zm-detail}

We summarize the essential steps in evaluating the direct nonvalence
zero-mode integral in Eq.~\eqref{eq:331}. After the
Pauli--Villars decomposition, it is sufficient to consider the generic
three-denominator contribution
\begin{equation}
\label{eq:Cgeneric}
I^{\rm Z.M.}_{ab}
\equiv
i\lim_{\alpha\to0^+}
\int_{\rm nv}\frac{d^4k}{(2\pi)^4}
\frac{
S^\perp_{\rm Z.M.}\cdot{\bm q}_\perp
}{
{\bm q}_\perp^2\,N_{\bar q}N_aN_b
},
\end{equation}
where $N_a$ and $N_b$ are the denominators introduced in
Appendix~A, with $a\in\{m_1,\Lambda_1\}$ and
$b\in\{m_2,\Lambda_2\}$.

In the nonvalence region, we introduce $x=\alpha z$ with
$0<z<1$ and take the upper-half-plane pole associated with $N_a$,
at which
$p_1^-=p^-_{1{\rm on}}(a)
=(a^2+{\bm p}_{1\perp}^2)/(\alpha zP^+)$.
With the shifted transverse momentum
${\bm\ell}_\perp={\bm p}_{1\perp}-z{\bm q}_\perp$,
the remaining denominators have the leading endpoint behavior
\begin{align}
\left.N_{\bar q}\right|_{N_a=0}
&\simeq
-P^+p^-_{1{\rm on}}(a),
\label{eq:zm-Nspectator}\\
\left.N_b\right|_{N_a=0}
&\simeq
-\frac{1}{z}
\left(
{\bm\ell}_\perp^2+B_{ab}
\right),
\label{eq:zm-Nfinal}
\end{align}
where
$B_{ab}=z(1-z){\bm q}_\perp^2+(1-z)a^2+zb^2$.
Here, $\simeq$ denotes equality at leading order in the endpoint
limit $\alpha\to0^+$.

Using the zero-mode numerator in Eq.~\eqref{eq:330}, its contraction
with the transverse momentum transfer becomes
\begin{align}
\left.
\frac{
S^\perp_{\rm Z.M.}\cdot{\bm q}_\perp
}{
{\bm q}_\perp^2
}
\right|_{N_a=0}
&\simeq
2P^+p^-_{1{\rm on}}(a)
\left[
\frac{
2{\bm\ell}_\perp\cdot{\bm q}_\perp
}{
{\bm q}_\perp^2
}
+
2z-1
\right].
\label{eq:zm-numerator-app}
\end{align}
The term linear in ${\bm\ell}_\perp$ vanishes under transverse
angular integration. After including the pole residue and using
$dx=\alpha\,dz$, the endpoint limit gives
\begin{equation}
\label{eq:Iab-zm-transverse}
I_{ab}^{\rm Z.M.}
=
\frac{1}{8\pi^3}
\int_0^1dz\,(2z-1)
\int d^2{\bm\ell}_\perp\,
\frac{1}{
{\bm\ell}_\perp^2+B_{ab}
}.
\end{equation}

The transverse integral for an individual Pauli--Villars term is
logarithmically divergent. In transverse dimensional regularization,
\begin{equation}
\left.
\int d^2{\bm\ell}_\perp\,
\frac{1}{
{\bm\ell}_\perp^2+B
}
\right|_{\rm DR}
=
\pi
\left[
{\rm div.}
-
\ln\left(\frac{B}{\mu^2}\right)
\right],
\end{equation}
where ${\rm div.}$ denotes a mass-independent divergent constant.
Consequently,
\begin{equation}
\label{eq:Iab-zm-final}
I_{ab}^{\rm Z.M.}
=
\frac{1}{8\pi^2}
\int_0^1dz\,(2z-1)
\left[
{\rm div.}
-
\ln\left(\frac{B_{ab}}{\mu^2}\right)
\right].
\end{equation}

The zero-mode contribution in Eq.~\eqref{eq:331} is obtained from the
alternating Pauli--Villars combination
\begin{equation}
\label{eq:C-PV-combination}
f_-^{\rm Z.M.}(q^2)
=
\Npv
\left[
I^{\rm Z.M.}_{m_1m_2}
-
I^{\rm Z.M.}_{m_1\Lambda_2}
-
I^{\rm Z.M.}_{\Lambda_1m_2}
+
I^{\rm Z.M.}_{\Lambda_1\Lambda_2}
\right].
\end{equation}
The mass-independent divergent constant and the dependence on the
regularization scale $\mu$ cancel in this combination.
Substituting Eq.~\eqref{eq:Iab-zm-final} into
Eq.~\eqref{eq:C-PV-combination} reproduces the direct zero-mode
result in Eq.~\eqref{eq:fmzm0}.

\section{Radial independence of the current-component relation}
\label{app:radial-identity}

Changes of variables from LF internal momenta to constituent
invariant masses are discussed in
Refs.~\cite{Melikhov1996Spectral,Melikhov2002Dispersion,
DesplanquesDong2008Translations,DesplanquesDong2009UnequalMasses}.
Here we use such invariant-mass variables only to expose the internal
orientation cancellation of the present BT overlap. The radial class
is
\begin{equation}
\phi_i
=
\sqrt{\frac{\partial k_{zi}}{\partial x}}\,
u_i(\rho_i),
\qquad
\rho_i
=
{\bm k}_{i\perp}^{\,2}+k_{zi}^2
=
\frac{\lambda(s_i,m_i^2,m_{\bar q}^2)}{4s_i},
\qquad
s_1=M_0^2,\quad s_2=M_0'^2,
\label{eq:radial-class}
\end{equation}
with the same pseudoscalar spin coupling and LF Jacobian as in the
main text. The real functions $u_i$ include their normalizations and
are otherwise unrestricted, provided the reduced overlaps converge.
Here $\lambda(z,u,v)=(z-u-v)^2-4uv$.

For $t=q^2<0$, let $\kappa=\sqrt{-t}$, $a=m_1$, $b=m_2$,
$c=m_{\bar q}$, and $v=1-x$. At fixed $(s_1,s_2)$, the remaining
$x$ dependence parametrizes an internal orientation. Defining
$d=s_1-s_2-\kappa^2=P^+q_{\rm on}^-$ and
$\alpha_\kappa=a^2-b^2-\kappa^2$, the allowed interval is determined by
\begin{align}
p(x)\equiv k_y^2
&=
-Lx^2+Bx-C
=
L(x_+-x)(x-x_-),
\nonumber\\
L
&=
s_1+\frac{d^2}{4\kappa^2},
\qquad
B
=
s_1+a^2-c^2
+\frac{\alpha_\kappa d}{2\kappa^2},
\qquad
C
=
a^2+\frac{\alpha_\kappa^2}{4\kappa^2},
\label{eq:invariant-orientation-variables}
\end{align}
where $0<x_-<x_+<1$.

For the radial class in Eq.~\eqref{eq:radial-class}, the combinations
$G_i(s_i)=xv\,\phi_i/D_i$, with
$D_i=\sqrt{\mathcal A_i^2+{\bm k}_{i\perp}^{\,2}}$, reduce to
functions of $s_i$ alone. Including the transformed measure and the
kernels in Eqs.~\eqref{eq:helicity-NF} and
\eqref{eq:helicity-HF}, the residual for $d\ne0$ becomes
\begin{equation}
\mathscr R(t)
=
-\frac{1}{32\pi^3\kappa}
\int_{\Omega_t}ds_1\,ds_2\,
\frac{G_1(s_1)G_2(s_2)\mathcal T}{d}
\int_{x_-}^{x_+}dx\,
\frac{2Lx-B}{\sqrt{p(x)}},
\qquad
\mathcal T
=
s_1+s_2+\kappa^2-2(a-c)(b-c),
\label{eq:radial-total-derivative}
\end{equation}
where $\Omega_t$ denotes the kinematically allowed invariant-mass
domain. All dependence on the radial functions lies outside the
orientation integral. Since $p'(x)=-2Lx+B$ and
$p(x_\pm)=0$,
\begin{equation}
\int_{x_-}^{x_+}
\frac{2Lx-B}{\sqrt{p(x)}}\,dx
=
-2\left[\sqrt{p(x)}\right]_{x_-}^{x_+}
=
0.
\label{eq:orientation-boundary-zero}
\end{equation}
It follows that $\mathscr R(t)=0$ independently of the detailed
radial shape within the convergent class defined in
Eq.~\eqref{eq:radial-class}.

The apparent $1/d$ singularity in
Eq.~\eqref{eq:radial-total-derivative} does not obstruct the
$d\to0$ limit. For $d\ne0$, pairing $x$ with $2x_c-x$, where
$x_c=B/(2L)$, cancels the orientation-odd contribution before this
limit is taken. The orientation-integrated projection therefore has
a finite continuation to $d=0$, although the unintegrated kernel is
singular there. Numerical calculations with power-law radial
functions $u_i\propto(1+\rho_i/\beta_i^2)^{-n}$ for $n=2,3$, using
the same LF Jacobian and pseudoscalar spin coupling, provide an
independent check of this cancellation.

In the numerical evaluation of Fig.~\ref{fig2} for $t<0$, the
minus-current overlap is evaluated in the original
$(x,{\bm k}_\perp)$ variables with the $x$ integration performed first
at fixed real ${\bm k}_\perp$. With ${\bm q}_\perp=(\kappa,0)$, the
denominator $xP^+q_{\rm on}^-=m_1^2-m_2^2-\kappa^2-2\kappa k_x$ does not
depend on $x$ and vanishes on the plane $k_x=k_x^*$, where
$k_x^*=(m_1^2-m_2^2-\kappa^2)/(2\kappa)$. On this plane $d=0$, so that
$s_2=s_1-\kappa^2$, $L=s_1$, and $B=s_1+m_1^2-m_{\bar q}^2$. Writing
$s_1=A_1/x+C_1/(1-x)$, with $A_1={\bm k}_\perp^2+m_1^2$ and
$C_1={\bm k}_\perp^2+m_{\bar q}^2$ held fixed, one finds
$x(1-x)\,ds_1/dx=2Lx-B$. The $x$-integrated numerator of the minus
overlap on this plane therefore becomes
$-\int_0^1dx\,G_1(s_1)G_2(s_1-\kappa^2)\mathcal T(s_1)\,ds_1/dx=0$.
Since $s_1(x)$ is real and tends to infinity at both $x\to0$ and
$x\to1$, the two branches traverse the same interval of $s_1$ in
opposite directions and cancel. For the Gaussian wave functions used
here, the $x$-integrated numerator is smooth in $k_x$, so its zero at
$k_x=k_x^*$ makes the singularity removable. The Gaussian damping
ensures absolute convergence of the remaining transverse integral.

\section{Regularity of the direct continuation}
\label{app:direct-continuation}

This appendix verifies the regularity and convergence of the full
Gaussian overlap integrands on the timelike contour of
Eq.~\eqref{eq:continued-contour} for $0<t\le t_{\rm max}$ in the
physical $D\to K$ region. The denominators, square-root branches,
and possible poles are examined below, followed by the finite
$t=0$ limits and numerical convergence checks.

\subsection{Regularity on the timelike contour}
\label{app:timelike-regularity}

On the contour of Eq.~\eqref{eq:continued-contour}, we write
$v=1-x$, $\alpha_1=\eta$, and $\alpha_2=1-\eta$, with
$\eta=m_1/(m_1+m_2)$. For this choice of $\eta$,
Eq.~\eqref{eq:physical-continuation-condition}, which is satisfied
for the $D\to K$ parameters used here, gives
$\alpha_i^2t_{\rm max}<m_i^2$ for $i=1,2$.
Throughout this subsection, $0<t\leq t_{\rm max}$ and the
integration variables $(x,{\bm\ell}_\perp)$ are real.

Polynomial factors of the integrand, such as the helicity numerators,
remain finite on the contour and need no discussion. A singularity or
branch point can arise only from the spin-factor square roots, the free
invariant masses in denominators, the Jacobians, and the vertex
functions, all of which depend on $t$ through ${\bm k}_{i\perp}^2$, and
from the minus-current denominator treated at the end of this
subsection. On the contour the quark transverse momenta are
${\bm k}_{1\perp}={\bm\ell}_\perp-\alpha_1v{\bm q}_\perp$ and
${\bm k}_{2\perp}={\bm\ell}_\perp+\alpha_2v{\bm q}_\perp$ with
${\bm q}_\perp=(i\sqrt t,0)$. Only their $x$ components,
$k_{i,x}=\ell_x\mp i\alpha_iv\sqrt t$, are complex.
To check that the spin-factor radicands and $M_{0i}^2$ remain
in the right half-plane, we first bound
${\rm Re}\,{\bm k}_{i\perp}^2$ from below. Since
${\bm k}_{i\perp}^2={\bm\ell}_\perp^2-\alpha_i^2v^2t
\mp2i\alpha_iv\sqrt t\,\ell_x$,
\begin{equation}
{\rm Re}\,{\bm k}_{i\perp}^2
={\bm\ell}_\perp^2-\alpha_i^2v^2t
\ge{\bm\ell}_\perp^2-\alpha_i^2v^2t_{\rm max}.
\label{eq:basic-estimate}
\end{equation}
At fixed $x$ and ${\bm\ell}_\perp$, the downward shift is
largest at zero recoil. Since $\alpha_i^2t_{\rm max}<m_i^2$,
its magnitude is strictly smaller than $v^2m_i^2$ for $0<x<1$.

\textit{Spin-factor square roots.} With
$\mathcal A_i=vm_i+xm_{\bar q}$, so that
$\mathcal A_i^2\ge v^2m_i^2+x^2m_{\bar q}^2$,
Eq.~\eqref{eq:basic-estimate} gives
\begin{equation}
{\rm Re}(\mathcal A_i^2+{\bm k}_{i\perp}^2)
\ge v^2\big(m_i^2-\alpha_i^2t_{\rm max}\big)
+x^2m_{\bar q}^2>0.
\label{eq:spin-roots}
\end{equation}
These arguments therefore neither vanish nor reach the negative real
axis, and the square roots are taken on the principal branch.

\textit{Free masses and Jacobians.} Inserting
Eq.~\eqref{eq:basic-estimate} into
$M_{0i}^2=(m_i^2+{\bm k}_{i\perp}^2)/x
+(m_{\bar q}^2+{\bm k}_{i\perp}^2)/v$
gives
\begin{equation}
{\rm Re}\,M_{0i}^2
=\frac{{\bm\ell}_\perp^2}{xv}
+\frac{m_i^2-\alpha_i^2t}{x}
+\frac{m_{\bar q}^2}{v}
+\alpha_i^2t
\ge m_i^2+m_{\bar q}^2
+2m_{\bar q}\sqrt{m_i^2-\alpha_i^2t_{\rm max}}
>\big|m_i^2-m_{\bar q}^2\big|.
\label{eq:free-mass-bound}
\end{equation}
The combination $(m_i^2-\alpha_i^2t)/x+\alpha_i^2t$ decreases
monotonically with $t$, since its derivative is
$-\alpha_i^2(1-x)/x\le0$, and is therefore bounded below by its value
at $t=t_{\rm max}$. The first inequality then follows from minimizing
over $x$. Hence $M_{0i}^2$ stays in the right half-plane and never
reaches the branch cut of the square root along the negative real axis.
The principal square root $M_{0i}=\sqrt{M_{0i}^2}$ is therefore finite
and nonzero along the contour and continuously approaches the positive
free invariant mass as $t\to0$. Moreover $|\Delta_i/M_{0i}^2|<1$ with
$\Delta_i=m_i^2-m_{\bar q}^2$, so the Jacobian factor
$1-\Delta_i^2/M_{0i}^4$ in Eq.~\eqref{eq:radial-jacobian-label} has a positive real part. 
We therefore take the principal square root of this factor.
Since ${\rm Re}\,M_{0i}>0$, multiplying this root by the principal
square root of $M_{0i}/(4xv)$ gives the branch of
$\sqrt{\partial k_{zi}/\partial x}$ in Eq.~\eqref{eq:9} that is positive
at $t=0$.

\textit{Gaussian factor and endpoints.} The Gaussian exponent
contains
$\vec k_i^{\,2}=M_{0i}^2/4-(m_i^2+m_{\bar q}^2)/2
+\Delta_i^2/(4M_{0i}^2)$.
The last term has a nonnegative real part, so
Eq.~\eqref{eq:free-mass-bound} gives the uniform bound
\begin{equation}
{\rm Re}\,\vec k_i^{\,2}
\ge\frac14\left[
\frac{{\bm\ell}_\perp^2}{xv}
+\frac{m_i^2-\alpha_i^2t_{\rm max}}{x}
+\frac{m_{\bar q}^2}{v}
\right]
-\frac{m_i^2+m_{\bar q}^2}{2}.
\label{eq:gaussian-bound}
\end{equation}
This bound is independent of $t$, so the Gaussian factors decrease
exponentially at large ${\bm\ell}_\perp^2$ and at both longitudinal
endpoints, uniformly for $0\le t\le t_{\rm max}$. The Gaussian damping
dominates the algebraic growth of the spin, Jacobian, and numerator
factors. The current-projector denominators are considered next.

\textit{Denominators and poles.} The plus- and transverse-current
projector denominators involve $\mathcal P^+>0$ and
${\bm q}_\perp^2=-t\ne0$ for $t>0$. In the physical $D\to K$ interval,
the pBT minus-current denominator also satisfies
$q^-=(M^2-M'^2+t)/P^+>0$. The fBT minus kernel in
Eq.~\eqref{eq:Kon-} contains instead the constituent factor
$q_{\rm on}^-$. On the contour,
\begin{equation}
xP^+q_{\rm on}^-
=m_1^2-m_2^2+[1-2\eta(1-x)]\,t-2i\ell_x\sqrt t.
\label{eq:continued-minus-denominator}
\end{equation}
The coefficient $1-2\eta(1-x)$ increases with $x$ from $1-2\eta$ to
$1$. For $\eta>1/2$, that is, for $m_1>m_2$, the real part of
Eq.~\eqref{eq:continued-minus-denominator} is therefore bounded below by
\begin{equation}
B_-\equiv m_1^2-m_2^2+(1-2\eta)t_{\rm max}
=(m_1-m_2)\left[
m_1+m_2-\frac{t_{\rm max}}{m_1+m_2}
\right]>0,
\label{eq:minus-denominator-bound}
\end{equation}
where the last inequality is the condition in
Eq.~\eqref{eq:physical-continuation-condition}. For $D\to K$,
$B_-\simeq1.9~\mathrm{GeV}^2$. Thus
$|xP^+q_{\rm on}^-|\ge B_->0$ on the whole contour, and the
minus-current denominator has no zero. For $m_1>m_2$, the regularity
of this denominator is implied by the wave-function analyticity bound.

Together with the Gaussian damping, these bounds yield a
$t$-independent integrable bound for the full integrand on every closed
interval $t_0\le t\le t_{\rm max}$ with $t_0>0$. The overlaps therefore
converge absolutely for each $t>0$ and, by dominated convergence, are
continuous throughout $0<t\le t_{\rm max}$. The transverse- and
minus-current extractions can be evaluated directly up to zero recoil.
The restriction $t_0>0$ keeps the explicit inverse powers of
${\bm q}_\perp^2=-t$ in the transverse-current kernels bounded, and the
endpoint $t=0$ is treated separately below.
The bounds on the spin denominators, free masses, Jacobians, and
minus-current denominator depend only on the constituent masses and
$t_{\rm max}$. The absence of singularities in the radial functions
themselves is a property of the Gaussian realization. Other radial
wave functions, such as the power-law forms of
Appendix~\ref{app:radial-identity}, require a separate analysis of
their own singularities on the continued contour.

At $t=0$ the contour reduces to real momenta. The complete transverse
and mapped Jaus kernels are then evaluated through their finite
angularly integrated limits rather than by pointwise substitution into
expressions containing explicit inverse powers of ${\bm q}_\perp^2$.
The fBT minus kernel contains no such factor.

\subsection{Numerical verification}

At the physical $q^2$ values on the 33-point grid of
Sec.~\ref{sec:numerical}, including both kinematic endpoints,
the independently evaluated transverse and fBT minus-current
extractions of $f_-$ agree to within an absolute difference of
about $2.5\times10^{-12}$ at quadrature order 120. The endpoint
values in Table~\ref{tab:endpoints} are stable under changes of the
quadrature order among 120, 160, and 200. The smoothness across $t=0$
was also checked by comparing $F(h)$ with $3F(0)-3F(-h)+F(-2h)$ for
$F=f_+$ and $f_-^{(\perp)}$. The differences decrease by a factor close
to eight when $h$ is halved, about $2\times10^{-7}$ at
$h=0.02~\mathrm{GeV}^2$, consistent with an $O(h^3)$ Taylor
remainder near $t=0$.
The timelike curves and endpoint values in Fig.~\ref{fig2} and
Table~\ref{tab:endpoints} are obtained from direct overlap integrations
without a form-factor parametrization or extrapolation.

\bibliographystyle{apsrev4-2}
\bibliography{references}
\end{document}